\pdfoutput=1  
\documentclass{new_tlp} 
\usepackage{amsmath}
\usepackage{amssymb}
\usepackage{mathptmx}
\usepackage{hyperref}
\usepackage[titletoc]{appendix}
\usepackage{yfonts}
\usepackage{listings}
\usepackage{proof} 
\title{In Praise of Impredicativity: A Contribution to the Formalisation of Meta-Programming}
  \author[Fran\c{c}ois Bry]
         {FRAN\c{C}OIS BRY\\
         Institute for Informatics, Ludwig-Maximilian University of Munich, Germany\\
         \email{bry@lmu.de}
         }

\jdate{November 2018}

\pagerange{\pageref{firstpage}--\pageref{lastpage}}
\doi{S1471068401001193} 
\newtheorem{definition}{Definition}[section]
\newtheorem{proposition}{Proposition}[section]
\newcommand{\rmodels}{\ensuremath{\mathrel{=\!\!\!|}}} 
\begin{document}

\label{firstpage}

\maketitle  

\begin{abstract}
Processing programs as data is one of the successes of functional and logic programming. Higher-order functions, as program-pro\-cess\-ing programs are called in functional programming, and meta-pro\-grams, as they are called in logic programming, are widespread declarative programming techniques. In logic programming, there is a gap between the meta-programming practice and its theory: The formalisations of meta-programming do not explicitly address its impredicativity and are not fully adequate. This article aims at overcoming this unsatisfactory situation by discussing the relevance of impredicativity to meta-progra\-mming, by revisiting former formalisations of meta-programming and by defining Reflective Predicate Logic, a conservative extension of first-order logic, which provides a simple formalisation of meta-programming. \\
{\it Under consideration for publication in Theory and Practice of Logic Programming}
\end{abstract}


\tableofcontents

\section{Introduction}\label{sec:Introduction}

Processing programs as data is one of the successes of functional and logic programming. Indeed, in most functional and logic languages, programs are standard data structures which release programmers writing program-processing programs from explicitly coding or importing  parsers. The following program, in which upper case characters are variables, specifies beliefs of Ann and Bill using the programming style called meta-programming:
\begin{lstlisting}
   believes(ann, itRains)
   believes(ann, itIsWet $\leftarrow$ itRains)
   believes(bill, X) $\leftarrow$ believes(ann, X)
\end{lstlisting}
This program's intended meaning is that Ann  believes that it rains, Ann believes that it is wet when it rains, and Bill believes everything Ann believes. This program is a meta-program because its second fact
\begin{lstlisting}
   believes(ann, itIsWet $\leftarrow$ itRains)
\end{lstlisting}
includes a clause: 
\begin{lstlisting}
   itIsWet $\leftarrow$ itRains
\end{lstlisting}
This fact violates the syntax of classical predicate logic that requires that a fact is formed from a predicate, like \lstinline{believes}, and a list of terms like \lstinline{ann}  but unlike the clause \lstinline{itIsWet $\leftarrow$} \lstinline{itRains}. Indeed, in classical predicate logic a clause is a formula, not a term.

Examples referring to beliefs and trust are given in this article because they are intuitive. However, this article does not address how to specify belief and trust systems but instead how to formalise meta-programming, a technique using which such systems can be specified. 

While most logics, 
especially classical predicate logic, prescribe a strict distinction between terms and formulas, meta-programming is based upon disregarding this distinction. 
Both Prolog and most formalisations of meta-progra\-mming
pay a tribute to this dictate of classical logic: They require to code a clause like
\begin{lstlisting}
   itIsWet $\leftarrow$ itRains
\end{lstlisting} 
as a compound term like 
\begin{lstlisting}
   cl(itIsWet, itRains)
\end{lstlisting}
or as an atomic term commonly denoted (using a so-called ``quotation'' $\ulcorner$ . $\urcorner$) as follows 
\begin{lstlisting}
   $\ulcorner$itIsWet $\leftarrow$ itRains$\urcorner$
\end{lstlisting}
when it occurs within a fact, expressing the second clause above in one of the following forms: 
\begin{lstlisting}
   believes(ann, cl(itIsWet, itRains))
   believes(ann, $\ulcorner$itIsWet $\leftarrow$ itRains$\urcorner$))
\end{lstlisting}

Such encodings or quotations are not necessary. Atomic and compound formulas can be treated as terms, as HiLog \cite{hilog} and
 Ambivalent Logic \cite{vademecum-1,vademecum-2} have shown. An expression such as 
\begin{lstlisting}
   likes(ann, bill)
\end{lstlisting}
(with the  intended meaning that Ann likes Bill) is built up from the three symbols \lstinline{likes}, \lstinline{ann} and \lstinline{bill} that all three can be used for forming nested HiLog, Ambivalent Logic and Reflective Predicate Logic expressions such as 
\begin{lstlisting}
   likes(ann, likes(bill, ann))
\end{lstlisting}
(with the intended meaning that Ann likes that Bill likes her). As a consequence, the Wise Man Puzzle suggested in the article \cite{wise-man-puzzle-1} as a benchmark for testing the expressive power and naturalness of knowledge representation formalisms can be expressed in HiLog, Ambivalent Logic and Reflective Predicate Logic exactly as it is expressed in the article \cite{wise-man-puzzle-2}.

\medskip

Like Ambivalent Logic \cite{vademecum-1,vademecum-2}, but unlike HiLog \cite{hilog}, Reflective Predicate Logic also allows expressions such as 
\begin{lstlisting}
    (loves $\land$ trusts)(ann, bill)
\end{lstlisting}
that can be defined by 
\begin{lstlisting}
    (loves $\land$ trusts)(X, Y) $\leftarrow$ loves(X, Y) $\land$ trusts(X, Y)\end{lstlisting}
or more generally by 
\begin{lstlisting}
    (P1 $\land$ P2)(X, Y) $\leftarrow$ P1(X, Y) $\land$ P2(X, Y)
\end{lstlisting}
and expressions such as 
\begin{lstlisting}
    likes(ann, (bill $\land$ charlie))
\end{lstlisting}     
that can be defined by:
\begin{lstlisting}
    P(X, (Y $\land$ Z)) $\leftarrow$ P(X, Y) $\land$ P(X, Z)
\end{lstlisting} 
Even more general expressions like the following are possible in Ambivalent Logic and Reflective Predicate Logic: 
\begin{lstlisting}
    ($\forall$ T trust(T) $\Rightarrow$ T)(ann, bill)
\end{lstlisting} 
or, in a  program syntax with implicit universal quantification
\begin{lstlisting}
    (T $\leftarrow$ trust(T))(ann, bill)
 \end{lstlisting} 
with the intended meaning that Ann trusts Bill, expressed as \lstinline{T(ann, bill)}, in all forms of trust specified by the meta-predicate \lstinline{trust}. If there are finitely many forms of trust, that is, if \lstinline{trust(T)} holds for finitely many values of \lstinline{T}, then this intended meaning can be expressed by the following rule that relies on negation as failure: 
\begin{lstlisting}
(T $\leftarrow$ trust(T))(X, Y) $\leftarrow$ not (trust(T) $\land$ not T(X, Y))
\end{lstlisting} 
The expression \lstinline{(T $\leftarrow$ trust(T))(ann, bill)} can also be proven in the manner of Gerhard Gentzen's Natural Deduction \cite{natural-deduction-1,natural-deduction-2,natural-deduction-3} by first assuming that \lstinline{trust(t)} holds for some surrogate \lstinline{t} form of trust that does not occur anywhere in the program, then proving \lstinline{t(ann, bill)} and finally discarding (or, as it is called ``discharging'') the  assumption \lstinline{trust(t)}. This second approach to proving 
\begin{lstlisting}
   (T $\leftarrow$ trust(T))(ann, bill)
\end{lstlisting}
is, in contrast to the first approach mentioned above, applicable if there are infinitely many forms of trust.

Even though Prolog's syntax does not allow compound predicate expressions such as 
\begin{lstlisting}
    (loves $\land$ trusts)
    (T $\leftarrow$ trust(T))
\end{lstlisting}
such expressions make sense. 

Reflective Predicate Logic has, in contrast to HiLog and Ambivalent Logic, an unconventional representation of variables. Its syntax adopts the  paradigm ``quantification makes variables''. Thanks to this paradigm, one can construct from the expression \lstinline{p(a, b)} in which \lstinline{a} and \lstinline{b} do not serve as variables, the expression \lstinline{$\forall$ a p(a, b)} in which \lstinline{a} serves as a variable and \lstinline{b} does not serve as variable. 
The paradigm ``quantification makes variables'' makes it easy to generate from an expression \lstinline{t(ann, bill)} a quantified expression \lstinline{$\forall$ t t(ann, bill)} which, as observed above, is needed in proving implications in the manner of  Natural Deduction  \cite{natural-deduction-1,natural-deduction-2,natural-deduction-3}. 
The paradigm ``quantification makes variables'' eases meta-programming as the following example shows. 
The formula  
\begin{lstlisting}
(believes(charlie,itRains) $\land$ believes(charlie,$\neg$itRains))
\end{lstlisting} 
(with the intended meaning that Charlie believes both, that it rains and that it does not rain) can easily be used in generating the (arguably reasonable) assertion

\begin{lstlisting}
($\exists$ itRains 
 (believes(charlie,itRains)$\land$believes(charlie,$\neg$itRains)))
 $\Rightarrow$ $\forall$ X believes(charlie, X)
\end{lstlisting} 
(with the intended meaning that if Charlie believes something and its negation, then Char\-lie believes everything) and also 

\begin{lstlisting}
$\forall$ charlie ($\exists$ itRains 
 ((believes(charlie,itRains)$\land$believes(charlie,$\neg$itRains)))
 $\Rightarrow$ $\forall$ X believes(charlie, X)
\end{lstlisting} 
(with the intended meaning that everyone believing something and its negation believes everything). 

A price to pay for the  paradigm ``quantification makes variables'' is that, in contrast to the widespread logic programming practice, universal quantifications can no longer be kept implicit. This is arguably a low price to pay since explicit universal quantifications are beneficial to program readability and amount to variable declarations that, since ALGOL 58 \cite{algol58-1,algol58-2} are considered a highly desirable feature of programming languages. Furthermore, explicit quantifications make system predicates like Prolog's \lstinline{var/1} that do not have a declarative semantics replaceable by declarative syntax checks because the presence of explicit universal quantifications distinguishes non-instantiated from instantiated variables. 
Another consequence of the  paradigm ``quantification makes variables'' is that Reflective Predicate Logic has no open formulas. This is, however, not a restriction, since open formulas have no expressivity in their own and serve only as components of closed formulas. It is even an advantage: Without open formulas, models are simpler to define. 

Reflective Predicate Logic can be seen as a late realisation, or rehabilitation, of Frege's logic \cite{begriffsschrift,Grundgesetze-der-Arithmetik-I,Grundgesetze-der-Arithmetik-II}. Except for the representation of variables, the syntax of Reflective Predicate Logic is a systematisation of the syntax of Frege's logic (see Section \ref{sec:Related-Work}). Therefore, the model theory given below can be seen as a model theory for Frege's logic. The name ``Frege's logic'' would have been given to the logic of this article if not for Frege's anti-democratic and anti-Semitic views. 

This article is structured as follows: 
Section \ref{sec:Introduction} is this introduction. 
Section \ref{sec:Requirements-to-Formalisations-of-Meta-Programming} considers a few meta-programs that motivate requirements to formalisations of meta-programming. 
Section \ref{sec:Related-Work} reports on related work. 
Section \ref{sec:Predicativity-Impredicativity} recalls why predicativity has been sought for and why impredicative atoms are acceptable. 
Section \ref{sec:Syntax} defines the syntax of Reflective Predicate Logic that allows impredicative atoms under the paradigm ``quantification makes variables''. 
Section \ref{sec:Paradoxes} discusses expressing the Barber and Russell's Paradoxes in Reflective Predicate Logic. 
Section \ref{sec:Variant-Rectification} gives a variant test for Reflective Predicate Logic expressions that is needed for the model theory of Reflective Predicate Logic. 
Section \ref{sec:Model-Theory} defines a model theory for Reflective Predicate Logic. 
Section \ref{sec:Symbol-Overloading} paves the way to the following section by recalling that symbols can be overloaded (as it is called in programming) in classical predicate logic languages. 
Section \ref{sec:Conservative-Extension} shows that Reflective Predicate Logic is a conservative extension of first-order logic. 
Section \ref{sec:Conclusion} concludes the article by discussing its contributions and giving perspectives for further work. 
A brief introduction into Frege's logic is given in an appendix. 

\medskip

The main contributions of this article are as follows: 
\begin{enumerate}
   \item A discussion of how Prolog-style meta-programming relates to Frege's logic, type theory, impredicativity, and Russell's Paradox of self-reflection. 
   \item A formalisation of meta-programming which is simple and a conservative extension of first-order logic. 
\item A model theory realising Frege's initial intuition that impredicative, or reflective, predicates can be accommodated in a predicate logic. 
\end{enumerate}

\section{Requirements to Formalisations of Meta-Programming}\label{sec:Requirements-to-Formalisations-of-Meta-Programming}

This section introduces two Prolog meta-programs so as to stress some aspects of meta-programm\-ing. 
The first meta-program is the well-known program \lstinline{maplist} \cite{The-Art-of-Prolog,The-Craft-Of-Prolog}: 
\begin{lstlisting}
   maplist(_, [], []).
   maplist(P, [X|Xs], [Y|Ys]) :-
      call(P, X, Y),
      maplist(P, Xs, Ys).
\end{lstlisting}
The third argument of \lstinline{maplist} is the list obtained by applying the first argument of \lstinline{maplist}, a binary predicate \lstinline{P}, to each element of the second argument of \lstinline{maplist} which is expected to be a list. If \lstinline{twice/2} is defined as: 
\begin{lstlisting}
   twice(X, Y) :- Y is 2 * X. 
\end{lstlisting}
then \lstinline{maplist(twice, [0,1,2], [0,2,4])} holds. 
\lstinline{maplist} is a meta-program because of the Prolog expression \lstinline{call(P, X, Y)} which builds from bindings of the variables \lstinline{P} and \lstinline{X} like \lstinline{P = twice} and \lstinline{X = 1} a fact like \lstinline{twice(1, Y)} and evaluates it. 
If the program \lstinline{maplist} would be seen as a set of classical predicate logic clauses, then \lstinline{call(P, X, Y)} would be expressed as  \lstinline{P(X, Y)}, the variable \lstinline{P} would be a second-order variable because it ranges over predicate symbols (like \lstinline{twice}) and the other variables would be first-order variables because they range over first-order terms (like integers or lists of integers).
The semantics of the meta-program \lstinline{maplist} can be conveyed by an infinite set of ground atoms such as: 
\begin{lstlisting}
   maplist(twice, [0,1,2], [0,2,4])
   maplist(length, [[a], [b,c]], [1,2])
   maplist(reverse, [[a,b], [c,d,e]], [[b,a], [e,d,c]])
\end{lstlisting}
The semantics of many, but not all, meta-programs can be similarly conveyed by ground atoms. 

The second example is a meta-program the semantics of which is not appropriately conveyed by ground atoms: 

\begin{lstlisting}
   studyProgram(mathematics).
   studyProgram(computing).
   
   syllabus(mathematics, logic).
   syllabus(computing,   logic). 
   syllabus(computing,   compilers).
   
   enrolled(student(anna), mathematics).
   enrolled(student(ben),  computing).

   attends(anna, logic).
   attends(ben,  compilers). 
   
   student(X) :- enrolled(student(X), _)
   
   course(C) :- syllabus(_, C).

   forall(R, F) :- not (R, not F).
\end{lstlisting}
The facts specify two study programs, their syllabi, the enrolments of students in study programs and the courses' attendance. The clauses with heads \lstinline{student(S)} and \lstinline{course(C)} extract the students' names and courses' titles respectively from the enrolments and syllabi. The clause with head \lstinline{forall(R, F)} serves to check properties like whether all mathematics students attend the course on logic: 
\begin{lstlisting}
   forall(enrolled(student(S), mathematics), 
                  attends(S, logic)
   )
\end{lstlisting}
or whether all students attend all the courses listed in their study programs' syllabi: 
\begin{lstlisting}
   forall(enrolled(student(S), P), 
                  forall(syllabus(P, C), attends(S, C)
                  )
   )
\end{lstlisting}
\lstinline{forall} is a meta-program because during its evaluation its arguments \lstinline{R} and \lstinline{F} are themselves evaluated. In the clause defining \lstinline{forall}, the left occurrences of \lstinline{R} and \lstinline{F} correspond to classical logic terms, while the right occurrences of the same variables correspond to classical logic formulas. 
\lstinline{forall} is a well-known Prolog meta-program which implements the failure-driven loop already mentioned in the introduction. It is representative of reflection in meta-programming, that is, the expression and the processing of formulas in meta-programs. The evaluation of for example 
\begin{lstlisting}
   forall(p(X), q(X))
\end{lstlisting}  
consists in a search for an instance of  \lstinline{p(X)} without corresponding instance of \lstinline{q(X)}. If the search fails, then the evaluation succeeds. Thus, the evaluation of 
\begin{lstlisting}
   forall(p(X), q(X))
\end{lstlisting} 
can be conveyed in classical logic by the formula: 
\begin{lstlisting}
   $\neg$$\exists$X (p(X) $\land$ $\neg$q(X))
\end{lstlisting}  
which, in classical logic, is logically equivalent to:  
\begin{lstlisting}
   $\forall$X (p(X) $\Rightarrow$ q(X))
\end{lstlisting}  
As a consequence, the instance of the clause defining the predicate \lstinline{forall}: 
\begin{lstlisting}
   forall(p(X), q(X)) :- not (p(X), not q(X)).
\end{lstlisting}  
corresponds to the classical logic formula 
\begin{lstlisting}  
   ($\forall$X (p(X) $\Rightarrow$ q(X)) $\Rightarrow$ $\forall$Y forall(p(Y), q(Y)))
\end{lstlisting}  
(in which, \lstinline{forall} is, like in the Prolog program, a predicate and \lstinline{X} and \lstinline{Y} are distinct variables) but does not correspond to the universal closure of the aforementioned clause instance: 
\begin{lstlisting}
   $\forall$X (p(X) $\Rightarrow$ q(X)) $\Rightarrow$ forall(p(X), q(X)))
\end{lstlisting} 
(in which only one variable occurs). 
As a consequence, ground instances of the meta-program
\begin{lstlisting}
   forall(R, F) :- not (R, not F).
\end{lstlisting}
like 
\begin{lstlisting}
    forall(p(a), q(a))
\end{lstlisting}
do not convey that meta-progam's semantics. Non-ground instances like 
\begin{lstlisting}
    forall(p(X), q(X))
\end{lstlisting}
are necessary to properly convey the semantics of \lstinline{forall}. 

In this, \lstinline{forall} is not a rare exception. Another example is the  meta-program \lstinline{hasSingVar(C)} that checks whether a clause \lstinline{C} contains at least one singleton variable. Its semantics cannot be described by ground instances like:
\begin{lstlisting}  
   hasSingVar(p(a) :- q(a, b))
\end{lstlisting}
Indeed, such a ground instance does not make sense. In contrast, a non-ground expression 
\begin{lstlisting} 
   hasSingVar(p(X) :- q(X, Y))
\end{lstlisting}
does convey the semantics of the meta-program \lstinline{hasSingVar(C)}. More generally, the semantics of many reflective meta-programs, of many meta-programs performing program analyses and of meta-programs generating improved program versions cannot be properly expressed by the ground expressions of standard Herbrand interpretations. 

A non-ground expression like the aforementioned one used for conveying a meta-program's semantics does not stand for the set of its ground instances. Thus, an explicit quantification like in 
\begin{lstlisting} 
   hasSingVar($\forall$ X $\forall$ Y p(X) :- q(X, Y))
\end{lstlisting}
better conveys the semantics of meta-programs. Conveying meta-programs' semantics with such non-ground expressions generalises Herbrand interpretations. 

Summing up, three aspects of meta-programming have been stressed in this section:  
\begin{enumerate}
   \item {\it Self-reflective predicates:} Some meta-predicates are self-reflective in the sense that they can occur within their own arguments like \lstinline{forall} in:
            \begin{lstlisting}
forall(enrolled(student(S), P), 
               forall(syllabus(P, C), attends(S, C)
               )
)
\end{lstlisting}
   \item {\it Confounding of object and meta-variables:} Some meta-programs contain occurrences of a same variable where classical logic expec ts a term and where classical logic expects a formula or a predicate like in:
             \begin{lstlisting}
forall(R, F) :- not (R, not F)
\end{lstlisting}
   \item {\it Need for generalised Herbrand interpretations:} The semantics of some meta-programs like \lstinline{forall} and \lstinline{hasSingVar(C)} is not properly conveyed by the ground atoms of standard Herbrand interpretations. It is appropriately conveyed by generalised Herbrand interpretations specified by non-ground and quantified expressions. 
\end{enumerate}
In the following section, the adquacy of formalisations of meta-programming is assessed by referring to the aforementioned three aspects of meta-programming. 

\section{Related Work}\label{sec:Related-Work}

This article relates to the many formalisations of meta-programming that have been proposed. These formalisations are of three kinds: 
\begin{itemize}
   \item Formalisations interpreting meta-programs as higher-order theories
   \item Formalisations interpreting meta-programs as first-order theories
   \item Formalisations interpreting meta-programs as theories in non-classical logics
\end{itemize}
The formalisations of meta-programming all relate to type theory. Therefore type theory is discussed below before the formalisations of meta-programming. As usual, the phrase ``type theory'' is used to refer either to the research field devoted to type theories or to a given type theory. This article also relates to reflection in computing, knowledge representation and logic and, because of examples it mentions, logics of knowledge and belief. 

\paragraph{Meta-programming} has been considered since the early days of logic programming. It is discussed in the article  \cite{logic-for-problem-solving,amalgamating}. Meta-programming in Prolog is addressed among others in the works \cite{clocksin-mellish,The-Art-of-Prolog,The-Craft-Of-Prolog}. However, the standard formalisation of logic programming \cite{Foundation-of-Logic-Programming} does not cover meta-programming. The articles \cite{Barklund-1995-Metaprogramming-in-Logic,Costantini-Meta-Reasoning-Survey-2002} survey meta-programming. 

Prolog's extremely permissive approach to meta-programming goes back to a fruitful disregard by Alain Colmerauer and Philippe Roussel, Prolog's designers, of the relationship between meta-programming, higher-order logic, impredicativity, and type theory and to a time at which the undecidability of unification in second-order logic \cite{undecidability-second-order-unification} and third-order logic \cite{undecidability-third-order-unification} as well as Damas-Hindley-Milner type systems \cite{type-systems-1,type-systems-2,type-systems-3,type-systems-4} were unknown or not widely known. Prolog's permissive approach to meta-programming is very useful in practice, as the following examples demonstrate: 
\begin{itemize}
   \item A unary predicate ranging over all unary predicates (including itself) can be used for (static or dynamic) type checking. 
   \item A predicate occurring in an argument of itself can be used for applying an optimisation to the very predicate specifying this optimisation.
   \item Formulas occurring in places where classical predicate logic expect terms are useful as it is shown in the articles \cite{logic-for-problem-solving,wise-man-puzzle-2}, in the introduction, and in the previous section. 
\end{itemize} 

\paragraph{Type theory.} The first type theory, or theory of types, was developed by Bertrand Russell as a correction of Gottlob Frege's  Logic \cite{begriffsschrift,Grundgesetze-der-Arithmetik-I,Grundgesetze-der-Arithmetik-II}, the archetype of classical predicate logic. Frege's logic is discussed below in Sections \ref{sec:Predicativity-Impredicativity} and  \ref{sec:Paradoxes} and its salient aspects are recalled in an appendix. Frege's logic is reflective in the sense that a predicate can apply to a formula or to a predicate, including itself. In Frege's logic, a unary predicate \lstinline{r} holding of all unary predicates that apply to themselves can be defined as follows:
\begin{lstlisting}
   $\forall$x (r(x) $\Leftrightarrow$ x(x))
\end{lstlisting}
As a consequence, Russell's Paradox \cite{link-russell-paradox}, which is discussed in more details below in Section \ref{sec:Paradoxes}, can be expressed in Frege's logic. In Frege's logic, some predicates cannot be interpreted as sets as the following set of atomic formulas illustrates in which \lstinline{a} and \lstinline{b} are individual constants and \lstinline{p} and \lstinline{q} are predicates: 
\begin{lstlisting}
   {p(a), p(b), q(a), q(q)}
\end{lstlisting}
If the constants \lstinline{a} and \lstinline{b} are interpreted as $1$ and $2$ respectively, then the predicate \lstinline{p} can be interpreted as the set $\{1, 2\}$ but the predicate \lstinline{q} cannot be interpreted as a set because a  set cannot be one of its own elements. Russell developed a type theory so as to avoid ``impredicative atoms'' like \lstinline{q(q)} and to preclude paradoxes of self-reflexivity like the paradox bearing his name. 

Russell successively developed various type theories before publishing with Alfred North Whitehead in Principia Mathematica \cite{Principia-Mathematica} the ``Ramified Theory of Types''. Leon Chwistek and Frank P. Ramsey  later simplified the Ramified Theory of Types yielding the theory now known as the ``Simple Type Theory'' or ``Theory of Simple Types'' \cite{Chwistek-Simple-Type-Theory-Polish,Ramsey-Simple-Type-Theory}. Aiming at avoiding paradoxes of the untyped lambda calculus \cite{church-untyped-lambda-calculus}, Alonzo Church re-expressed the Simple Type Theory as a theory which is now commonly called ``Church's Simply Typed Lambda Calculus'', or $\lambda^{\to}$, \cite{church-lambda-calculus,typed-lambda-calculus}, a typed variation   (with a single type constructor, $\to$, for function types) of the untyped lambda calculus.

Every type theory requires that every symbol, among others every variable, and every compound expression have a single type like first-order term, first-order predicate, first-order formula, second-order predicate, second-order formula, etc. In other words, a type theory imposes a strong typing. This strong typing is Russell's treatment of impredicative atoms like \lstinline{q(q)} and of paradoxes of self-reflexivity among others Russell's Paradox: It precludes them. Endowed with, and modified by, the strong typing of a type theory, Frege's logic became classical predicate logic. 

Referring to the typing policy of classical predicate logic as a ``strong typing'' is an anachronism. Indeed that denomination has been introduced  only in 1974 in the article  \cite{strong-typing} where it is defined as follows: ``Whenever an object is passed from a calling function to a called function, its type must be compatible with the type declared in the called function.'' This definition perfectly describes the requirement of a type theory if ``object'' is understood as ``argument'' and ``function'' as ``predicate or function'', which justifies the anachronism. 

A strong typing in the manner of, but different from, the type theory of classical predicate logic has been shown later to be useful for preventing programming errors \cite{cardelli-wegner}. This has resulted in ``type systems'' that assign properties to program constructs \cite{Types-and-Programming-Languages,Type-Systems}. These properties and type systems depart from the ``orders'' or ``types'' of the aforementioned type theories. 


Remarkable achievements in type theory (among others the Curry-Howard isomorphism, inductive types and dependent types) 
have given type theories an aura of indispensability. This article shows that type theories can be dispensed with: It gives a model theory to a systematisation of Frege's logic, Reflective Predicate Logic, a logic without type theory, yielding a simple and intuitive formalisation of Prolog-style meta-programming. 

\paragraph{Formalisations interpreting meta-programs as higher-order theories.}
Some logic programming languages, most prominently $\lambda$Prolog \cite{Lambda-Prolog,lambda-prolog-book-2012}, Elf \cite{Elf} and Twelf \cite{Twelf}, are formalised in classical higher-order predicate logics, syntactical restrictions ensuring necessary properties like the decidability of unification. 

$\lambda$Prolog is based on the Simply Typed Lambda Calculus \cite{church-lambda-calculus,typed-lambda-calculus}. Elf \cite{Elf} and Twelf \cite{Twelf} have been designed conforming to the Edinburgh Logical Framework LF \cite{LF}, a predicative language for a uniform representation of the syntax, the inference rules, and the proofs of predicative logics. LF is based on intuitionistic logic \cite{intuitionistic-logic} and on the Simply Typed Lambda Calculus \cite{church-lambda-calculus,typed-lambda-calculus}. Surprisingly, LF represents object variables by meta-variables \cite[p.~145]{LF}: 
\begin{quote}
``$[\ldots]$ substitution in the logical system is encoded as substitution in LF (which relies on the identification of object-logic variables with the variables of LF).''
\end{quote}
This identification, or confounding, of object and meta-variables is, in fact, precluded by the Simply Typed Lambda Calculus on which LF is based (as it is precluded by any other type theory). This article can be seen as a justification for LF's confounding of object and meta-variables. 

William W.\ Wadge has proposed in the article \cite{wadge-extensionality} a fragment of higher-order Horn logic called ``definitional programs'' as a meta-programming language ``based on Church's Simple Theory of Types''  \cite{wadge-extensionality}, that is, the Simply Typed Lambda Calculus or $\lambda^{\to}$ \cite{church-lambda-calculus,typed-lambda-calculus}. The particularity of Wadge's language is that its semantics fulfils a condition called ``extensionality'' that makes predicates with a same extension interchangeable. Extensionality is justified in the article \cite{wadge-extensionality} with the following example: 
 \begin{lstlisting}
   p(a).
   q(a).
   phi(p). 
\end{lstlisting}
The predicates \lstinline{p} and \lstinline{q} are interpreted in the article \cite{wadge-extensionality,extensional-holp} as first-order predicates. 
\lstinline{phi} is a meta-program interpreted in the article \cite{wadge-extensionality,extensional-holp} as a second-order predicate which applies to the predicate \lstinline{p}. The argument given in the article \cite{wadge-extensionality} for justifying extensionality is that \lstinline{phi(q)} should hold like \lstinline{phi(p)} because \lstinline{p} and \lstinline{q} have the same extension. Marc Bezem has given in the article \cite{bezem-good-programs} under the name of ``good programs'' a first decidable sufficient condition for extensionality as an improvement over Wadge's ``definitional programs'', and under the name of ``hoapata programs'' \cite{bezem-extensionality}  a second decidable sufficient  condition for extensionality as an improvement of the ``good programs''. 

Extensionality is a questionable requirement because it abstracts out the algorithmic aspects of programs as the following example shows. Replace in the aforementioned example \lstinline{p} by the tail-recursive program \lstinline{reverse} and \lstinline{q} be the non-tail-recursive program \lstinline{naivereverse} \cite{The-Art-of-Prolog,The-Craft-Of-Prolog}. Recall that \lstinline{reverse} and \lstinline{naivereverse} have the same extension.  Assume that the meta-predicate \lstinline{phi} is a versioning predicate \lstinline{final} distinguishing final implementations from less efficient preliminary versions. Since \lstinline{final(reverse)} holds,  \lstinline{final(naivereverse)} must, by virtue of extensionality, also hold. Clearly, this is not desirable. 

All formalisations of meta-programs as higher-order theories adhere to the strong typing of classical predicate logic: They strictly distinguish between terms and formulas and therefore cannot express meta-programs like 
\begin{lstlisting}
    forall(R, F) :- not (R, not F). 
    r(X) :- X(X).
\end{lstlisting}
in which some variables stand for both a term and a formula or a predicate. All formalisations interpreting meta-programs as higher-order theories preclude self-reflective predicates and the confounding of object and meta-variables of Prolog-style meta-programming. The formalisations interpreting meta-programs as higher-order theories do not provide generalised Herbrand interpretations specified by non-ground and quantified expressions.  Thus, the formalisations of meta-programs as higher-order theories do not fulfil the requirements stressed at the end of Section \ref{sec:Requirements-to-Formalisations-of-Meta-Programming}.

\paragraph{Formalisations interpreting meta-programs as first-order theories}
are based on reifying first-order formulas by encoding them as first-order terms. The advantage of the approach over formalisations of meta-programming in higher-order logic is that it makes possible  reflective, including self-reflective, formulas. Its drawbacks are the encodings that make programs complicated and less intuitive than their Prolog-style counterparts. The encodings upon which formalisations of meta-programming are based are called "naming relations" or "naming schemes". A formalisation of meta-programming in first-order logic is largely, if not fully, characterised by its naming relation.

A naming relation refers to two first-order languages, an object language ${\cal L}_O$ and a meta-language ${\cal L}_M$, for encoding terms, especially variables, and formulas of ${\cal L}_O$ as terms of ${\cal L}_M$. Some naming relations encode all terms and all formulas of ${\cal L}_O$, others encode only some of the terms and formulas of ${\cal L}_O$. Naming relations typically encode variables of ${\cal L}_O$ as variable-free, or ground, terms of ${\cal L}_M$ so as to avoid the confounding of object and meta-variables of Prolog-style meta-programming which, as it is recalled above, is precluded by the strong typing of classical predicate logic. 
Three kinds of naming relations have been considered: 
\begin{itemize}
   \item A {\it structure-hiding naming relation} encodes the structure of an object formula as a meta-term the structure of which does not reflect that of the formula. 
   \item A {\it quotation} encodes an object formula $F$ without conveying its structure, typically as a meta-language constant usually denoted ${\ulcorner}{F}{\urcorner}$.
   \item A {\it structure-preserving naming relation} encodes a formula as a meta-term the structure of which reflects that of the formula. 
\end{itemize}

Kurt G\"{o}del and Alfred Tarski  first used naming relations for encoding self-reflective formulas in first-order logic that, because of classical predicate logic's strong typing, cannot be directly expressed in that logic. G\"{o}del used the following naming relation \cite{Goedel-Incompleteness-Theorems-1931}: If $F$ is the object language formula represented as $s_1 s_2 \ldots s_n$ where the $s_i$ are non-negative integers representing the object language symbols, then $F$ is encoded as the non-negative integer $G(F) = p_1^{s_1} \times p_2^{s_2} \ldots \times p_n^{s_n}$, the G\"{o}del number of $F$, where $p_1 = 2, p_2 = 3, p_3 = 5, \ldots$ is the ordered sequence of the prime numbers. Because of the Unique-Prime-Factorization theorem, a formula $F$ can be reconstructed from its G\"{o}del number $G(F)$, that is, the naming relation is structure-hiding. Shortly later, Tarski used quotations for expressing his ``Schema $T$'' \cite{Tarski-Theory-of-Truth-1935}, that is, the requirement that every theory of truth have a truth predicate $T$ fulfilling $(T({\ulcorner}{F}{\urcorner}) \Leftrightarrow F)$ for all formulas $F$. 

The first reference to a naming relation in logic programming is probably the article \cite{amalgamating}. In that article, provability of an object language $L$ is expressed in a meta-language $M$ with a predicate $Demo$ implementing SLD resolution \cite{sld-resolution}. $Demo$ is inspired from G\"{o}del's predicate $Bew$ \cite{Goedel-Incompleteness-Theorems-1931}, short for Beweis, that is, proof in German. 

Under the naming relation of \cite{amalgamating}, the formulas of the object language $L$ are encoded as terms of the meta-language $M$. The atom \lstinline{p(X, bill)} is for example encoded as the term \lstinline{atom(pred(1),} \lstinline{[var(1), constant(212)])} in which \lstinline{pred(1)}, \lstinline{var(1)}, and \lstinline{constant(212)} are variable-free, or ground, terms of $M$. Thus, the naming relation considered in the article \cite{amalgamating} is structure-preserving. Following a widespread logic programming practice, in both $L$ and $M$ commas denote conjunctions, universal quantifications are implicit and existential quantifications are not used. As a consequence, the article \cite{amalgamating} can avoid to address that a naming relation in fact requires to encode in the meta-language the logical connectives and quantifiers of the object language. 

In the article \cite{amalgamating}, the amalgamation of an object language $L$ and a meta-language $M$ is defined as ``$L$ and $M$'', meaning $L \cup M$, equipped with:
\begin{itemize} 
   \item a naming relation which associates with every  expression of $L$ a variable-free term of $M$,
   \item a representation of $\vdash_L$ in $M$ by means of a predicate {\it Demo} specified in M as a theory $Pr$,
   \item{the ``linking rules'' 

            \vspace{0.3em}   
            \hspace{2em}\infer{A \vdash_L B}{Pr \vdash_M Demo(A', B')}\hspace{4em}\infer{Pr \vdash_M Demo(A', B')}{A \vdash_L B} 

           in which $F'$ denotes the encoding of $F$ under the naming relation. 
           }
\end{itemize}
The linking, or reflection or attachment, rules have been proposed in the article \cite{Weyhrauch-Prolegomena-1980}, a formalisation of meta-level reasoning. They correspond to the correctness and completeness of G\"{o}del's predicate $Bew$ \cite{Goedel-Incompleteness-Theorems-1931} with respect to provability. 
The article \cite{amalgamating} states that ``the amalgamation of $L$ and $M$ is a conservative extension in the sense that no sentence is provable in the amalgamation that is not in either $L$ or $M$.'' This statement disregards sentences of the amalgamation that may contain symbols of $L$ but not of $M$ as well as symbols of $M$ but not of $L$ and therefore that are sentences of neither $L$ nor $M$. 
 
The article \cite{amalgamating} further states that ``the amalgamation allows to have $L = M$, where the two languages are identical'' meaning that a language may contain an encoding of its own formulas as terms. 

The article \cite{Semantical-Properties-of-Encodings-1995} formalises naming relations as rewrite systems and investigates the expressivity of various naming relations. The article \cite{Harmelen-Definable-Naming-Relations-META92} argues that naming relations can encode, together with an object formula, pragmatic and semantic information resulting in a more efficient  (meta-language) version of the original formula. Naming relations have indeed been defined for achieving such ``compilations'' what explains their large number and, as a consequence, the large number of formalisations of meta-programming in first-order logic: metaProlog \cite{metaProlog-1,metaProlog-2}, MOL \cite{Eshghi-1986-MOL}, the language proposed by Barklund in the article \cite{Barklund-1989-What-is-a-meta-variable-in-Prolog}, Reflective Prolog \cite{Reflective-Prolog-1,Reflective-Prolog-2}, R-Prolog$^{*}$ \cite{r-prolog-star-1,r-prolog-star-2}, 'LOG (spoken ``quotelog'') \cite{quotelog}, G\"{o}del \cite{Goedel}, the language proposed by Higgins in the article \cite{higgins} and the generalisation of Reflective Prolog proposed in the article \cite{reflection-principles}. 

Most formalisations of meta-programming in first-order logic make use of naming relations that are structure-preserving because they are ``compositional'', that is, recursively defined on the expressions' structures, yielding names (or encodings) satisfying
\begin{center}
$˜\overline{f(g(a), b)} \approx \overline{f}( \overline{g}( \overline{a}),  \overline{b})$ 
\end{center}
where $\overline{e}$ denote the encoding of an expression $e$. The approximation $\approx$ cannot always be replaced by an equality because of cases like the aforementioned one \cite{amalgamating}: $\overline{p(t_1, t_2)} = atom(\overline{p}, [\overline{t_1}, \overline{t_2}]) \neq \overline{p}(\overline{t_1}, \overline{t_2})$. Some formalisations of meta-programming in first-order logic make use not only of structure-preserving naming relations but also of quotations giving compound expressions short ``names'' that, in general, are individual constants of the meta-language. Some other formalisations of meta-programming in first-order logic make use only of quotations. Note that the denomination ``quotation'' is often used in the meta-programming literature in the sense of ``structure-preserving naming relation'' instead of the aforementioned sense of encodings of formulas that hide the formulas' structures. 
Most formalisations of meta-programming in first-order logic are amalgamations of object and meta-languages in the sense of \cite{amalgamating} recalled above. 

metaProlog \cite{metaProlog-1,metaProlog-2} has two naming relations such that ``constants act as names of themselves. For non-constant items, metaProlog provides structural or non-structural names (and sometimes both), where the former are compound terms whose structure reflects the syntactic structure of the syntactic item they name.'' metaProlog has  ``metalevel names'' that are terms used for representing object programs. Object variables are represented in metaProlog as constants of the meta-language. metaProlog treats theories, or programs, as named entities that, thanks to the aforementioned naming relation can be referred to in a program. This makes possible that, in metaProlog, goals are proven in reference to a named theory and updates are formalised in logic. metaProlog has an explicit quantification that avoids problems in updating theories similar to those stressed above in Section \ref{sec:Requirements-to-Formalisations-of-Meta-Programming} while discussing the semantics of the meta-program \lstinline{forall}. metaProlog provides ``methods for moving between a name and the thing it names [$\ldots$] analogous to univ (\lstinline{=..}) of ordinary Prolog.'' Thus, in metaProlog, an object language expression can be obtained from its encoding as a meta-term. 

MOL \cite{Eshghi-1986-MOL} has a structure-preserving naming relation and an involved treatment of reflection through ``inheritance and scoping axioms'': An inheritance axiom  can be used to express that an object-level program $P$ contains the object-level program $Q$;  a scoping axiom is used to express that if a ground assertion can be proved from the meta-theory $M$, then this ground assertion is part of a ``description'' of an object-level program $P$, that is, that $M$ is a meta-theory for $P$. In MOL, an object language expression can be obtained from its encoding as a meta-term. MOL supports self-reflection called ``self-reference'' in the thesis \cite{Eshghi-1986-MOL}. This thesis shows how reflection and self-reflection can be used in meta-programming and for expressing ``non-floundering negation as failure'' and a ``declarative control [of program execution] without jeopardising the soundness of the interpreter.'' 

The language proposed by Barklund in the article \cite{Barklund-1989-What-is-a-meta-variable-in-Prolog} has a  naming relation for ``a naming of Prolog formulas and terms as Prolog terms'' built from reserved function symbols and constants ``to create a practical and logically appealing language for reasoning about terms, programs.'' In this language, an object language expression can be obtained from its encoding as a meta-term. 

Reflective Prolog \cite{Reflective-Prolog-1,Reflective-Prolog-2} has a naming relation, called ``quotation'' in the article, that for example encodes, or ``names'', the term \lstinline{f(a)} as \lstinline{function(functor($\{$f$\}$),} \lstinline{arity(1),}  \lstinline{args(["a"]))}. The article mentions that ``it is possible to build names of names of names, and so on'' but does not explain the use of this feature. Reflection is well supported by Reflective Prolog through an ``unquotation mechanism'' realised by ``a distinguished truth predicate which relates names to what is named.'' Reflective Prolog is based on an amalgamation of object and meta-languages that are disjoint in the sense that they have no symbols in common. As a consequence, ``language and metalanguage are amalgamated in a non-conservative extension'':  ``Statements are provable in the amalgamated language, that are provable neither in the language nor in the metalanguage alone.'' Reflective Prolog has an ``extended resolution procedure which automatically switches the context between levels'' which ``relieves the programmer from having to explicitly deal with control aspects of the inference process.'' Reflective Prolog's ``extended resolution is proved sound and complete with respect to the least reflective Herbrand model.''

R-Prolog$^{*}$ \cite{r-prolog-star-1,r-prolog-star-2} has a naming relation called ``quotation'' and denoted \`{a} la Lisp  \cite{LISP1.5-Manual,let-s-talk-lisp} with a single quote: If \lstinline{t = f(a, b)}, then \lstinline{'t =  'f('a, 'b)}. R-Prolog$^{*}$ furthermore has predicates $\uparrow$ (spoken ``up'') and $\downarrow$ (spoken ``down'') such that  \lstinline{$\uparrow$t $=$ 't} and  \lstinline{$\downarrow$'t $=$ t}. Thus, in R-Prolog$^{*}$, an object language expression can be obtained from its encoding as a meta-term. R-Prolog$^{*}$ reifies not only object terms and formulas but also substitutions so as to express much of the language's runtime system in its own meta-language. The semantics of R-Prolog$^{*}$ is based on an ``extended notion of interpretations and models'' that departs from the usual semantics of logic programs based on a fixpoint of the immediate consequence operator or Herbrand interpretations: ``Because computational reflection is a procedural notion, we cannot adopt the usual declarative semantics.''

'LOG \cite{quotelog} has two naming relations associating ``two different but related meta-representations with every syntactic object of the language, from characters to programs'', ``a constant name and a structured ground term, called the structural representation.'' This double naming relation generalises those of \cite{Martelli-Rossi-Enhancing-Prolog-1988,Rossi-Meta-programming-Facilities-1989,Rossi-Programs-as-Data-1992} that apply only to programs. 'LOG has an operator \lstinline{<=>} that relates the name and the structural representation of each syntactic object. 'LOG has no ``mechanism which would allow a meta-representation to be obtained from the object it denotes or vice versa.'' The article stresses that ``this differentiates (both in aims and in nature) our proposal from others, such as Reflective Prolog and R-Prolog$^{*}$, that, on the contrary, assume a reflection mechanism to be available, though not visible at the user level." 

G\"{o}del \cite{analysis-metaprograms,Goedel} is equipped with a strong typing system distinguishing object language from meta-language expressions and a naming relation encoding object language expressions as constants of the meta-language called ``ground representations.''  G\"{o}del's naming relation is thus a quotation. G\"{o}del makes the encoding of object language expressions by the naming relation, ``explicitly available to the programmer.'' G\"{o}del expresses much of the language's deduction system in its meta-language: It reifies among other the object language's provability. G\"{o}del has no mechanism for obtaining an object expression from its ``ground representation'', that is, its encoding in the meta-language. G\"{o}del is a rare case of a formalisation of meta-programming in first-order logic which is not an amalgamation of the object and meta-language in the sense of the article \cite{amalgamating} recalled above. 

Higgins has proposed in the article \cite{higgins} a language relying on two naming relations similar to those of 'LOG associating with every syntactic object a first-order ``primitive name'' and a first-order ``structured name'': ``We have names of symbols, terms, clauses, and sets of clauses.'' The procedural semantics of Higgins' language is ``a resolution rule and a meta-level to object-level reflection rule.'' In other words, the language's procedural semantics obtains an object language expression from its encoding as a meta-term. 

The articles \cite{Semantical-Properties-of-Encodings-1995,Semantical-Properties-of-SLD-Resolution-1995,reflection-principles} generalise Reflective Prolog \cite{Reflective-Prolog-1,Reflective-Prolog-2} into a metalanguage similar in spirit to LF \cite{LF}, Elf \cite{Elf}, and Twelf \cite{Twelf} which offers a sublanguage for expressing various kinds of naming relations (or encodings), primitives for handling naming relations, and a metalanguage of Horn clauses for expressing object-level inference rules. 
The article \cite{Semantical-Properties-of-Encodings-1995} considers eight previously proposed naming relations (or encodings) that are compositional in the sense that the name of a compound expression is built from its components' names. The article submits that naming relations should be compositional and shows that compositional naming relations can be expressed as rewrite systems provided the systems preserve truth (and falsity), are confluent and are terminating. 
The article \cite{Semantical-Properties-of-SLD-Resolution-1995} extends the generalisation of Reflective Prolog to SLD Resolution. The article  \cite{reflection-principles} further specifies a fixpoint semantics for definite programs in the proposed metalanguage, establishes the soundness and completeness of SLD resolution with respect to that fixpoint semantics and gives three examples of meta-programs one of which is a re-implementation of Reflective Prolog  \cite{Reflective-Prolog-1,Reflective-Prolog-2}. 

All formalisations interpreting meta-programs as first-order theories adhere to the strong typing of classical predicate logic: They strictly distinguish between terms and formulas. As a consequence, they cannot directly express meta-programs in which some variables stand for both a term and a formula or a predicate and self-reflective predicates or formulas like: 
\begin{lstlisting}
    forall(R, F) :- not (R, not F). 
    r(X) :- X(X).
\end{lstlisting}
However, through naming relations, that encode object expressions as meta-terms, and object variables as non-variable meta-terms, they can express such examples. 

Whether such (encoded) examples can be efficiently processed depends among other things on whether the encoding is invertible, that is, whether a decoding primitive is available that returns from a code the object language expression its encodes and whether such a decoding primitive can be efficiently implemented.
If the encoding is efficiently invertible, then adjustments to the deduction methods suffice to implement a language that handles meta-programs by relying on naming relations. metaProlog \cite{metaProlog-1,metaProlog-2}, MOL \cite{Eshghi-1986-MOL}, the language proposed by Barklund in the article \cite{Barklund-1989-What-is-a-meta-variable-in-Prolog}, Reflective Prolog \cite{Reflective-Prolog-1,Reflective-Prolog-2}, R-Prolog$^{*}$ \cite{r-prolog-star-1,r-prolog-star-2}, and the language proposed by Higgins in the article \cite{higgins} have efficiently invertible naming relations. 'LOG  \cite{quotelog} and G\"{o}del \cite{Goedel} do not have invertible naming relations. 

However, as many authors have observed, with a strong typing preventing the confounding of object and meta-var\-iables, much of the object language reasoning, among others unification, must be re-implemented in the meta-language \cite{De-Schreye-Martens-1992-sensible-least-herbrand,Kalsbeek-1993-Vanilla-Metainterpreter,meta-programming-fo-real-1993,Barklund-Costantini-1995-SLD-Resolution-with-Reflection,praise-ambivalent-syntax}. Even if such re-implementations are provided with a language, they are a burden to the programmers. Furthermore, such re-implementations result in significant losses in efficiency. Finally, such re-implementations are redundant since the object language's reasoning is a special case of the meta-language's reasoning. These observations have triggered a debate dubbed ``ground versus non-ground representations'' that revolves around the following question: If the theory requires that object and meta-languages expressions, especially variables, be distinguished, then why are well-working deduction systems possible that confound object and meta-variables? So far, the debate was not settled. This article shows that the logic in which meta-programming is formalised can be adapted to Prolog-style meta-programming giving a formal justification to the practice of confounding object and meta-variables in deduction systems. 

Summing up, the formalisations of meta-programs as first-order theories do fulfil the first of the three requirements mentioned in Section \ref{sec:Requirements-to-Formalisations-of-Meta-Programming}: Through encodings, they can express self-reflective predicates. However, they do not fulfil the second requirement of  Section \ref{sec:Requirements-to-Formalisations-of-Meta-Programming}: They cannot confound object and meta-variables.  Some of these formalisations do not, other partly fulfil the third requirement of  Section \ref{sec:Requirements-to-Formalisations-of-Meta-Programming}:
Those formalisations relying on non-ground encodings do not have generalised Herbrand interpretations specified by non-ground atoms while those formalisations relying on ground encodings can express (ground encodings of) such atoms. Generalised Herbrand interpretations with atoms including quantified formulas have not been considered by the authors of formalisations of meta-programs as first-order theories. 

\paragraph{Formalisations~ interpreting~ meta-programs~ as theories in non-classical logics}
adapt first-order logic to fit Prolog-style meta-programming by giving up type theory, which makes reflective, including self-reflective, formulas and the confounding of object and meta-variables possible. 
A predicate logic without type theory has already been defined at the end of the 19th century: Frege's logic \cite{begriffsschrift,Grundgesetze-der-Arithmetik-I,Grundgesetze-der-Arithmetik-II}, the precursor of classical predicate logic. A brief presentation of Frege's logic is given in an appendix. The following formula (see Section \ref{sec:Requirements-to-Formalisations-of-Meta-Programming} above) that confounds object and meta-variables  can be expressed in Frege's logic: 
\begin{lstlisting}
   $\forall$ x (r(x) $\Leftrightarrow$ x(x))
\end{lstlisting} 
Modified by the addition of a type theory, Frege's logic became classical predicate logic. Unsurprisingly, the formalisations of meta-programming in non-classical logics are closely related to Frege's logic. 

The syntax of Frege's logic has been defined before inductive definitions (and grammars as formalisms easing the expression of inductive definitions) were established (see Section \ref{sec:Predicativity-Impredicativity}). As a consequence, Frege did not fully formalise expressions resulting from replacing in an expression a subexpression by its definition according to his logic's Basic Law V (see the appendix which introduces into Frege's logic). Consider for example the following definition of \lstinline{p}: 
\begin{lstlisting}
   (p $\Leftrightarrow$ (q $\land$ r))
\end{lstlisting} 
Replacing \lstinline{p} by its definition in \lstinline{p(a)} results in the following expression with compound predicate: 
\begin{lstlisting}
   (q $\land$ r)(a)
\end{lstlisting} 
If \lstinline{q} is defined by: 
\begin{lstlisting}
   (q $\Leftrightarrow$ r(b, c))
\end{lstlisting} 
then replacing \lstinline{q} by its definition in \lstinline{q(a)} results in the following ``multiple application:'' 
\begin{lstlisting}
   (r(b, c))(a)
\end{lstlisting} 
``Multiple applications'' are known from currying. Currying, common in Functional Programming but rare in predicate logic, is used in automated deduction among other for implementing efficient term indexes \cite{Peter-Graf-Term-Indexing}. 
The aforementioned substitutions are not possible in first-order logic because of its strong typing that precludes the predicate definitions they result from. Indeed, since \lstinline{p},  \lstinline{q} and  \lstinline{r} are unary predicates, instead of for example \lstinline{(p $\Leftrightarrow$ (q $\land$ r))} first-order logic would require: 
\begin{lstlisting}
   $\forall$ x (p(x) $\Leftrightarrow$ (q(x) $\land$ r(x)))
\end{lstlisting} 
Frege's logic has a proof calculus but no model theory. It was only in 1930 that G\"{o}del would introduce in his doctoral thesis \cite{goedel-thesis-german} the concepts of interpretation and model in establishing the completeness of Frege's proof calculus for a fragment of Frege's logic, first-order logic. Below, Section \ref{sec:Syntax} gives Frege's logic the systematised syntax discussed above (under the paradigm ``quantifications make variables'') and Section \ref{sec:Model-Theory} a Herbrand-style model theory. This results in a simple and intuitive formalisation of Prolog-style meta-programming. 

HiLog \cite{hilog} was the first formalisation of meta-programming in a non-classical logic. At first, HiLog seems to be a formalisation in first-order logic because the article \cite{hilog} refers to a type theory and a naming relation. However, HiLog treats every symbol (except connectives and quantifiers) as a predicate and maps it to both 
\begin{itemize}
   \item an ``infinite tuple of functions'' over the universe, one function of arity $n$ for each $n \in \mathbb{N}$
   \item an ``infinite tuple of relations'' over the universe, one relation  of arity $n$ for each $n \in \mathbb{N}$
\end{itemize}
As a consequence, every HiLog expression built without connectives and quantifiers is, in the sense of first-order logic, both a term and an atomic formula. Since every expression being of all types amounts to no expressions being of any type and since a (standard) type theory assigns a single type to an expression, HiLog can be seen as a logic without (standard) type theory. In the article \cite{hilog}, a naming relation is used for encoding HiLog in first-order logic but not for encoding formulas as terms in HiLog: Reflection in HiLog is achieved without naming relation. As a consequence, HiLog can express meta-programs like the following (see Section \ref{sec:Requirements-to-Formalisations-of-Meta-Programming}) without resorting to a naming relation: 
\begin{lstlisting}
    forall(R, F) :- not (R, not F). 
    r(X) :- X(X).
\end{lstlisting}

HiLog's syntax is the Horn fragment with implicit universal quantifications of the aforementioned systematised syntax of Frege's logic. Thus, though reflection is possible in HiLog, it is limited to that fragment. The following statement (with the intended meaning that Ann believes that it rains and the sun shines) cannot be expressed in HiLog because HiLog does not allow connectives to occur within an atom:   
\begin{lstlisting}
   believes(ann, (itRains $\land$ theSunShines))
\end{lstlisting}
However, the article \cite{hilog} rightly claims that allowing connectives within atoms is a minimal extension. The first of the following statements cannot be expressed in HiLog because HiLog does not allow quantifiers within atoms while the second and the third of the following statements can be expressed in HiLog through skolemisation:  
\begin{lstlisting}
   believes(ann, $\exists$ x (logician(x) $\land$ loves(x, ann)))
   $\exists$ x (logician(x) $\land$ believes(ann, loves(x, ann)))
   $\exists$ x believes(ann, (logician(x) $\land$ loves(x, ann)))
\end{lstlisting}
Note the difference between the three aforementioned statements: The first expresses that Ann believes to be loved by a logician, the second
expresses the existence of a logician Ann believes to be loved by, the third  expresses the existence of an entity Ann believes to be a logician who loves her. In a world without logicians, the first statement can be fulfilled, the second not. In a world without enough entities, the number of which might be limited by the axioms considered, the first statement could be fulfilled and the third not. Such differences are significant in meta-programming, especially in meta-programs performing program analyses. The requirement in a module $A$ for another module $B$ should for example not be misinterpreted as stating the existence of a module $B$. The article \cite{hilog} states that ``encoding formulas with quantified variables would require introduction of lambda-abstraction which can be done but is out of the scope of this paper.'' It is correct that this can be done, though not through lambda-abstraction. It has been done in the articles  \cite{vademecum-1,vademecum-2} in a manner, however, which is not satisfying. An appropriate treatment of quantifiers within atoms is given below in Section \ref{sec:Model-Theory}. 
       
HiLog has a Herbrand model theory that specifies the semantics of the meta-program \lstinline{maplist} (see Section \ref{sec:Requirements-to-Formalisations-of-Meta-Programming})
as a set of ground facts like
\begin{lstlisting}
   maplist(twice, [0,1,2], [0,2,4])
\end{lstlisting}
However, HiLog's model theory is not satisfying for non-ground atoms like the following (see Section \ref{sec:Requirements-to-Formalisations-of-Meta-Programming}): 
\begin{lstlisting}
   forall(enrolled(student(S), P), 
                  forall(syllabus(P, C), attends(S, C)
                  )
   )
\end{lstlisting}
Indeed, HiLog's model theory interprets such a non-ground atom as a set of ground atoms. Recall the need for generalised Herbrand models specified by non-ground and quantified expressions stressed at the end of Section \ref{sec:Requirements-to-Formalisations-of-Meta-Programming}. 
The article \cite{hilog} claims that ``under 'reasonable' assumptions'' a HiLog language can be given a ``classical semantics.'' Clearly, this is only possible by equipping the language with a type theory that would keep apart terms from formulas. This would considerably restrict the language's meta-programming capability and, in fact, ruin HiLog's objectives. 
HiLog has a specific treatment of equality based on paramodulation \cite{paramodulation-1,paramodulation-2,paramodulation-3}. 
Finally, the article \cite{hilog} neither mentions that Russell's Paradox (see below Section \ref{sec:Paradoxes}) is expressible in HiLog nor refers to Frege's logic. 

Ambivalent Logic \cite{vademecum-1,vademecum-2} is a second formalisation of meta-programming in a non-classical logic. It is explicitly defined as a non-classical predicate logic that does not distinguish between terms and formulas. Thus, even though the articles \cite{vademecum-1,vademecum-2} do not mention type theory, they nonetheless specify a logic without type theory. The syntax of Ambivalent Logic is the aforementioned systematised syntax of Frege's logic (a few additional parentheses, necessary for disambiguation, are missing in its definition). 
Ambivalent Logic has a Herbrand-style model theory in which the meta-program \lstinline{maplist} (see Section \ref{sec:Requirements-to-Formalisations-of-Meta-Programming}) is expressed as a set ground facts like:
\begin{lstlisting}
   maplist(twice, [0,1,2], [0,2,4])
\end{lstlisting}
Ambivalent Logic can express meta-programs like the following (see Section \ref{sec:Requirements-to-Formalisations-of-Meta-Programming}) without resorting to a naming relation: 
\begin{lstlisting}
    forall(R, F) :- not (R, not F). 
    r(X) :- X(X).
\end{lstlisting}
The treatment in Ambivalent Logic's model theory of existentially quantified formulas, in contrast to that of HiLog, is not based on skolemisation. As a consequence, as stated in the article \cite{vademecum-2}: 
\begin{quote}
``$[\ldots]$  unlike in Hilog, both $\exists p \forall x (p(x) \leftrightarrow \neg q(x))$ and $\exists p \forall x(p(x) \leftrightarrow \exists u (q(u)(x))$ are valid in AL $[$Ambi\-valent Logic$]$.'' \cite[p.~55]{vademecum-2}
\end{quote}
In contrast to HiLog, Ambivalent Logic can express as follows that Ann believes to be loved by a logician: 
\begin{lstlisting}
   believes(ann, $\exists$ x (logician(x) $\land$ loves(x, ann)))
\end{lstlisting}
The treatment in Ambivalent Logic's model theory of non-ground atoms like:
\begin{lstlisting}
    forall(p(X), q(X))
    forall(p(Y), q(Y))
\end{lstlisting} 
or
\begin{lstlisting}
    believes(bill, $\forall$ X believes(ann, X))
    believes(bill, $\forall$ Y believes(ann, Y))
\end{lstlisting} 
is not satisfying. As pointed out in the articles \cite{vademecum-1,vademecum-2}, in a same Ambivalent Logic interpretation, the one variant expression of each example can be true and the other false: 
\begin{quote}
``$[\ldots]$ It should be noted that `similar' expressions like, for example, $\forall x.f(x)$ and $\forall y.f(y)$ constitute different, and unrelated, objects in the domains of models. That is, the truth values of the closed expressions $t(\forall x.f(x))$ and $t(\forall y.f(y))$ need not be the same.'' \cite[p.~38]{vademecum-2}
\end{quote}
In contrast to HiLog, Ambivalent Logic has no specific treatment of equality. Thus, an Ambivalent Logic theory can, like a classical predicate logic theory, have normal interpretations that interpret the equality predicate $=$ as the equality relation as well as other interpretations that interpret it as an equivalence relation which might be useful in logic programming or for knowledge representation. The article \cite{vademecum-2} discusses a property called ``opaqueness'' and states: 
\begin{quote}
``$[\ldots]$ the schema $\forall x \forall y.(x = y \rightarrow \phi(x) \leftrightarrow \phi(y))$, and also $\forall x \forall y \forall z.(x = y \rightarrow x(z) \leftrightarrow y(z))$ $[$hold in HiLog$]$. In contrast, in the context of AL $[$Ambivalent Logic$]$ we have a choice between validating the above schema or not, by either taking all of ET $[$the equality axioms$]$ as the equality theory, or restricting the equality to ET(I,II) $[$the equality axioms I and II$]$. Thus, unlike AL $[$Ambivalent Logic$]$, HiLog is not appropriate for intensional logics, where opaqueness is usually desirable.'' \cite[p.~55]{vademecum-2}
\end{quote}
Finally, the articles  \cite{vademecum-1,vademecum-2} do not mention that Russell's Paradox (see below Section \ref{sec:Paradoxes}) is expressible in Ambivalent Logic and do not refer to Frege's logic. 

Like Ambivalent Logic, Reflective Predicate Logic has a more expressible syntax than HiLog that allows quantifiers and connectives to appear within atoms. The syntaxes of Ambivalent Logic and Reflective Predicate Logic are systematisations of the syntax of Frege's logic that differ only in their representations of variables. The Herbrand-style model theory of Reflective Predicate Logic is similar to those of HiLog and Ambivalent Logic. It is more general than that of HiLog and corrects a serious deficiency of that of Ambivalent Logic by ensuring that in an interpretation, variant expressions are identically interpreted. Like the model theories of classical predicate logic and Ambivalent Logic, and unlike the model theory of HiLog, the model theory of Reflective Predicate Logic is not constrained to a specific treatment of equality.

HiLog, Ambivalent Logic and Reflective Predicate Logic fulfil the first two requirements to formalisations of meta-programming mentioned in Section \ref{sec:Requirements-to-Formalisations-of-Meta-Programming}: They can express self-reflective predicates and they confound object and meta-variables. HiLog does not fulfil the third requirement of Section \ref{sec:Requirements-to-Formalisations-of-Meta-Programming}: It has no generalised Herbrand interpretations specified by non-ground and quantified expressions.  Ambivalent Logic does fulfil this third requirement although in an unsatisfying manner. Reflective Predicate Logic corrects this deficiency of Ambivalent Logic. 

\paragraph{Reflection in computing, knowledge representation, and logic.}
A language is reflective if statements can be expressed in this language that refer to themselves or other statements of the same language.
Reflection is ubiquitous in computing: The program-as-data paradigm is a form of reflection, the von Neumann architecture is reflective, some programming languages are reflective, reflection is used for proving undecidability results, reflection is often needed in knowledge representation because introspective capabilities are often required from intelligent software and robots, etc.

Formalisations of meta-programming in first-order logic are closely related to reflection in knowledge representation \cite{Weyhrauch-Prolegomena-1980,Perlis-1985-Languages-with-Self-Reference-I,Aiello-1986-Representation-and-Use-of-Metaknowledge,Perlis-1988-Languages-with-Self-Reference-II,Perlis-1988-Meta-in-Logic,Harmelen-1989-Classification,Benjamin-1990-Manifesto}. 

 
\paragraph{Logics of knowledge and belief.}

The examples referring to the beliefs of agents given in this article could be expressed in a logic of knowledge and belief \cite{Hintikka-1962-Belief,Halpern-1985-Guide,Lismont-1994-Common-Belief,van-Ditmarsch-2015-Handbook-Epistemic-Logic}.

\section{Predicativity and Impredicativity}\label{sec:Predicativity-Impredicativity}

Predicativity \cite{predicativity} is an essential trait of classical logic atoms. Meta-programming makes use of impredicative atoms. This section recalls one of the reasons why impredicative atoms have been banned from classical predicate logic. 

Consider a property $P$ on the nodes of an undirected graph $G$ defined as follows: A node $n$ of $G$ has property $P$ if all its immediate neighbours have property $P$. This definition is not acceptable because it is ambiguous: It applies among others to  the property holding of no nodes and to the property holding of all nodes.  

At the beginning of the 20th century Henri Poincarr\'{e} and Bertrand Russell have proposed the Vicious Circle Principle  \cite{russell-vicious-circle-principle-1,russell-vicious-circle-principle-2} that forbids circular definitions, that is, definitions referring to the very concept they define like the above definition of property $P$. Russell called ``predicative'' definitions that adhere to the Vicious Circle Principle, ``impredicative'' definitions that violate it. Thus, the Vicious Circle Principle is the adhesion to predicativity and the rejection of impredicativity.

The Vicious Circle Principle, however, has a drawback: It forbids hereditary and, more generally, inductive definitions \cite{inductive-definitions} like the definition of the formulas of a logic, or of the programs of a programming language, or of the fixpoint of the immediate consequence operator of a definite logic program  \cite{fixpoint-semantics-1,fixpoint-semantics-2}. (Recall that the definition of a property $P$ is hereditary if it states that whenever a natural number $n$ has property $P$, so does $n + 1$. Recall that a definite logic program is a program the clauses of which contain no negative literals.)
The Vicious Circle Principle also rejects the definition sketched at the beginning of this section even though this definition makes sense as an inductive definition:
\begin{itemize}
  \item Base cases: A (possibly empty) set of nodes of $G$ is specified that have the property $P$.  
  \item  Induction case: If a node has the property $P$, then all its immediate neighbours have the property $P$. 
\end{itemize}
An inductively defined property (or set) is the smallest property (or set) that fulfils the base and induction cases of its definition \cite{inductive-definitions}. Thus, understood as an inductive definition, the definition sketched at the beginning of this section is that of the empty relation, that is, of the relation that holds of no nodes. 

Since the semantics of recursive functions and predicates is defined in terms of inductive definitions, the Vicious Circle Principle also implies the rejection of recursive functions and predicates. Clearly, such a rejection is not compatible with  programming. 
The Vicious Circle Principle further rejects definitions like the following that are widely accepted even though they are not inductive and they do not provide constructions of the entities they define: 
\begin{itemize}
  \item $y$ is the smallest element of an ordered set $S$ if and only if for all elements $x$ of $S$, $y$ is less than or equal to $x$, and $y$ is in $S$.
   \item The definition of the stable models of logic programs \cite{stable-model-semantics}.
\end{itemize}
Such definitions have in common that they quantify over domains the definitions of which refer to the entities being defined.  
Examples like the aforementioned led some mathematicians, most notably G\"{o}del, to object that impredicative definitions are acceptable provided the entities they refer to are clearly apprehensible \cite{Russells-Mathematical-Logic}. 
Nowadays, most logicians and mathematicians follow G\"{o}del and accept impredicative definitions of the following kinds \cite{inductive-definitions,predicativity}:
\begin{itemize}
  \item Inductive definitions.
  \item Impredicative definitions that characterise elements (like the smallest number in a set) of clearly apprehensible sets (including inductively defined sets).
\end{itemize}

The Vicious Circle Principle was the reason for Russell and Frege, whom Russell persuaded, to reject impredicative atoms built up from predicates that ``apply to themselves'', like a predicate expressing a set of all sets. Such predicates are expressible in Frege's logic  \cite{begriffsschrift,Grundgesetze-der-Arithmetik-I,Grundgesetze-der-Arithmetik-II}, the precursor of first-order logic. A predicate that ``applies to itself'' is the core of Russell's Paradox \cite{link-russell-paradox} (discussed below in Section \ref{sec:Paradoxes}).
The Vicious Circle Principle and paradoxes, among other the paradox bearing his name, motivated Russell to develop the Ramified Theory of Types \cite{ramified-theory-of-types} that precludes predicates ``applying to themselves''. 

Because classical logic adheres to a type theory, classical logic rejects reflective expressions like the following that are at the core of meta-programming:
\begin{lstlisting}
    believes(ann, itRains) 
    believes(ann, believes(bill, itRains))
\end{lstlisting}
where \lstinline{itRains} might be true or false, that is, amounts to a formula, not a term. 

In order to simplify the following argument, let us consider a unary predicate ``belief'' derived from the above definitions by disregarding who is holding a belief: 
\begin{lstlisting}
    belief(itRains) 
    belief(belief(itRains))
\end{lstlisting} 
On the one hand, the unary predicate \lstinline{belief} cannot be interpreted by a set $B$ because $B$ would have as elements the subset of all beliefs like \lstinline{itRains} that are believed to be believed. On the other hand, a set of closed atoms like the above two expressions perfectly gives a semantics to the unary predicate \lstinline{belief}. 

Relying in such a manner on a standard set of closed atoms for defining non-standard ``sets'', say ``collections'', like the  collection of beliefs in the above example, is the essence of  the model theory proposed below. The resulting impredicative definitions (in the example given above, the definition of beliefs) fulfils G\"{o}del's condition to be interpreted in reference to clearly apprehensible entities (in the example given above, a standard set of closed atoms). 



\section{Syntax of Reflective Predicate Logic}\label{sec:Syntax}

This section introduces ``expressions''  that amount to both the terms and the formulas of classical predicate logic languages. Except for the use of the paradigm ``quantification makes variables'' and a more careful parenthesising ensuring that expressions are non-ambiguous, the syntax given below is that of Ambivalent Logic \cite{vademecum-1,vademecum-2}. 

\begin{definition}[Symbols and Expressions]\label{def:expressions}
A Reflective Predicate Logic language $\cal L$  is defined by
   \begin{itemize}
      \item the logical symbols consisting of 
               \begin{itemize}
                   \item 
                   the connectives $\wedge$,  $\vee$, $\Rightarrow$, and $\neg$,
                   \item 
                   the quantifiers $\forall$ and $\exists$, 
                   \item
                   the parentheses ) and ( and the comma , .
               \end{itemize}
      \item at least one and at most finitely many non-logical symbols each of which is distinct from every logical symbol.
\end{itemize}
The expressions of a Reflective Predicate Logic language $\cal L$ and their outermost constructors are inductively defined as follows: 
   \begin{itemize}
      \item A non-logical symbol $s$ is an expression the outermost constructor of which is $s$ itself. 
     \item  If $E$ and $E_1, \ldots , E_n$ ($n \geq 1$) are expressions, then $E(E_1, \ldots, E_n)$ is an expression
               the outermost constructor of which is $E$.  
     \item  If $E$ is an expression, then $(\neg E)$ is an expression the outermost constructor of which is $\neg$. 
     \item  If $E_1$ and $E_2$ are expressions, then $(E_1 \land E_2)$, $(E_1 \lor E_2)$, $(E_1 \Rightarrow E_2)$ are
               expressions the outermost constructors of which are $\land$, $\lor$, and $\Rightarrow$ respectively. 
     \item If $E_1$  and $E_2$ are expressions, 
              then $(\forall E_1~  E_2)$ and $(\exists E_1~  E_2)$ are expressions the outermost constructors of which are $\forall$ and $\exists$ respectively. 
   \end{itemize}
A logical or non-atomic expression is an expression the outermost constructor of which is a connective or a quantifier. A non-logical or atomic expression, or atom, is an expression the outermost constructor of which is neither a connective nor a quantifier.
\end{definition}

%

The set of expressions of a Reflective Predicate Logic language is not empty since, by definition, the language has at least one non-logical symbol.

More parentheses are required by Definition \ref{def:expressions} than in classical predicate logic and than stated in the article \cite{vademecum-2}. This is necessary for distinguishing (well-formed) expressions such as $(\neg a)(b)$ and $(\neg a(b))$ or $(\forall x~ p(x))(a)$ and $(\forall x~ p(x)(a))$. Provided a few additional parentheses are added, first-order logic formulas are expressions in the sense of Definition \ref{def:expressions}, that is, the syntax given above is a conservative extension of the syntax of first-order logic. This issue is addressed in more detail below in Section \ref{sec:Conservative-Extension}.

The expressions of a Reflective Logic language can be proven non-ambi\-gu\-ous similarly as classical logic formulas are proven non-ambiguous. The subexpressions and proper subexpressions of an expression of a Reflective Predicate Logic language can be similarly defined as the sub-formulas and proper sub-formulas of a classical logic formula. Thus, an expression is a subexpression but not a proper subexpression of itself. 

The definition of the scope of a quantified variable in an expression generalises that of classical predicate logic: The scope of $E_1$ in ($\forall E_1~ F$) and in ($\exists E_1~ F$) is $F$ except those subexpressions of $F$ of the form ($\forall E_2~ G$) or ($\exists E_2~ G$) such that $E_2$ is a subexpression of $E_1$. 
Thus, in each of the following expressions, the inner quantified expression is not within the scope of the outer quantified expression: 
\begin{lstlisting}
   ( $\forall$ x     ( $\forall$ x     p(f(x), x)  ) )
   ( $\forall$ x     ( $\forall$ x     p(x(f), x)  ) )
   ( $\forall$ f(a)  ( $\forall$ f     p(f(a), f)  ) )
\end{lstlisting}
In each of the following expressions that can be obtained from each other by a variable renaming, the inner quantified expression is within the scope of the outer quantified expression.  
\begin{lstlisting}
   ( $\forall$ f  ( $\forall$ x     p(x,    f) ) )
   ( $\forall$ f  ( $\forall$ f(a) p(f(a), f)  ) )
\end{lstlisting}

Recall that a logical, or non-atomic, expression is an expression the outermost constructor of which is a negation, a connective, or a quantifier. Thus, 
\begin{lstlisting}
    (believes(ann, itRains) $\land$ believes(ann, itIsWet))
    ($\exists$ X believes(ann, X))
    ($\neg (\exists$ Y (believes(bill, Y))))
\end{lstlisting}
are logical expressions. Logical expressions correspond to first-order logic compound (that is, non-atomic) or quantified  formulas. 

Recall that an atomic expression, or atom, is an expression the outermost constructor of which is neither a connective nor a quantifier.
Thus, 
\begin{lstlisting}
    believes(ann, (itRains $\wedge$ itIsWet))
    believes(ann, ($\forall$ X believes(bill, X)))
    (believes $\wedge$ trusts)(ann, bill)
    ($\forall$ T (trust(T) $\Rightarrow$ T))(ann, bill)
 \end{lstlisting}
are atoms while 
\begin{lstlisting}
    (believes(ann, itRains) $\land$ believes(ann, itIsWet))
    ($\forall$ X believes(bill, X))
\end{lstlisting}
are non-atomic, or logical, expressions. Note that if $A_1$ and $A_2$ are atoms, then the (well-formed) expressions ($\forall A_1 A_2$) and ($\exists A_1 A_2$) are not atoms. Note also that an expression is either atomic (or non-logical) or non-atomic (or logical). 

Skolemisation can be specified as usual by adding additional non-logical symbols to the language.

For the sake of simplicity, the definition above assumes that every non-logical symbol has all arities. This reflects a widespread logic programming practice: Using \lstinline{p/2}, that is, \lstinline{p} with arity 2, in a Prolog program, for example, does not preclude using \lstinline{p/3}, that is, \lstinline{p} with arity 3, in the same program. Assuming that non-logical symbols of a Reflective Predicate Logic language have all arities is a convenience, not a necessity. The definition above can be refined to a less permissive definition as of non-logical symbols' arities. 

In contrast to the syntax of Ambivalent Logic given in the articles \cite{vademecum-1,vademecum-2}, the above definition does not distinguish between variables and constants. According to the above definition, quantifications make variables: 
\begin{itemize}
   \item \lstinline{likes(ann, bill)} contains no variables. In this expression, \lstinline{ann} and \lstinline{bill} serve as constants. 
   \item \lstinline{($\exists$ ann likes(ann, bill))}  means that there is someone who likes Bill. In this expression \lstinline{ann} serves as a variable and \lstinline{bill} as a constant. 
   \item \lstinline{($\forall$ bill ($\exists$ ann likes(ann, bill)))} means that everyone is liked by someone. In this expression \lstinline{ann} and \lstinline{bill} serve as a variables. 
\end{itemize}
A first advantage of the paradigm ``quantifications make variables'' is that every expression is closed. Indeed, a symbol which is not quantified such as \lstinline{x} in \lstinline{likes(x, bill)} is not a variable. This is not a restriction, since in logics with open formulas, open formulas serve only as components of closed formulas.
The paradigm ``quantifications make variables''  corresponds to the declarations of programming languages. The paradigm ``quantification makes variables'' is akin to lambda-abstraction. We give it an expressive denomination for avoiding referring to the lambda calculus our proposal does not build upon.

Explicit quantifications as introduced in Definition \ref{def:expressions} are not usual in logic progra\-mming. Combined with the paradigm ``quantifications makes variables'', they are useful for meta-progra\-mming because they make it easy to transform expressions. The expression \lstinline{likes(ann, bill)} for example can be abstracted into  
\begin{lstlisting}
    ($\exists$ likes likes(ann, bill))
 \end{lstlisting}
(meaning that Ann and Bill are in some relationship) and generalised as 
\begin{lstlisting}
  $(\forall$ likes likes(ann, bill))
 \end{lstlisting}
(meaning that Ann and Bill are in all possible relationships). Similarly, the expression
\begin{lstlisting}
    ((($\exists$ x p(x)) $\land$ ($\neg$ ($\exists$ x p(x)))) $\Rightarrow$ ($\forall$ G  G))
\end{lstlisting} 
(meaning that every expression follows from \lstinline{($\exists$ x p(x))} and its negation) can easily be generalised into 
\begin{lstlisting}
($\forall$ ($\exists$x p(x)) ((($\exists$x p(x)) $\land$ ($\neg$($\exists$x p(x)))) $\Rightarrow$ ($\forall$G G)))
\end{lstlisting} 
that is, after renaming \lstinline{F} the expression \lstinline{($\exists$ x p(x))} serving as variable,
\begin{lstlisting}
    ($\forall$ F ((F $\land$ ($\neg$ F)) $\Rightarrow$ ($\forall$ G G)))
\end{lstlisting} 
(meaning that every expression follows from an expression and its negation). 
 
\section{The Barber and Russell's Paradoxes in Reflective Predicate Logic}\label{sec:Paradoxes}

One of the reasons for the Vicious Circle Principle, that is, the rejection of impredicative definitions, was Russell's Paradox, a second-order variation of the first-order Barber Paradox. Both the Barber and Russell's paradoxes can be expressed in Reflective Predicate Logic. This section explains why this is not a problem.

Since Reflective Predicate Logic's syntax is a conservative extension of the syntax of first-order logic, a formulation of the Barber Paradox in first-order logic like the following is also a formulation of that paradox in Reflective Predicate Logic: 
\begin{lstlisting}
   man(barber) 
   ($\forall$y (man(y) $\Rightarrow$ (shaves(barber,y) $\Leftrightarrow$ ($\neg$shaves(y,y)))))
\end{lstlisting}
where, extending Definition \ref{def:expressions}, $(E_1 \Leftrightarrow E_2)$ is defined as a shorthand notation for $((E_1 \land E_2) \lor ((\neg E_1) \land (\neg E_2)))$. (This extension is a common manner to define the semantics of $\Leftrightarrow$ in classical logic.) 
The above expressions convey that the barber is a man shaving all men who do not shave themselves. The Barber Paradox is a mere inconsistency: The barber cannot exist because he would have both to shave himself and not to shave himself. The self-contradictory formula 
\begin{lstlisting}
   (shaves(barber,barber) $\Leftrightarrow$ ($\neg$shaves(barber,barber))) 
\end{lstlisting}
follows in first-order logic from the above specification of the Barber Paradox. 
A formula expressing the Barber Paradox is inconsistent with respect to the model theory defined in the section after next as it is in first-order logic.

The syntax of Section \ref{sec:Syntax} that does not distinguish between formulas and terms gives rise to (well-formed) expressions that are not expressible in first-order logic, and that, like the Barber Paradox, are inconsistent. One such expression is the following that expresses Russell's Paradox in both Frege's logic \cite{begriffsschrift,Grundgesetze-der-Arithmetik-I,Grundgesetze-der-Arithmetik-II} and in Reflective Predicate Logic (a brief introduction into Frege's logic is given in an appendix): 
\begin{lstlisting}
   ($\star$) ($\forall$x (e(x) $\Leftrightarrow$ ($\neg$x(x)))) 
\end{lstlisting}
Instantiating x with e in ($\star$) yields the self-contradictory expression 
\begin{lstlisting}
   (e(e) $\Leftrightarrow$ ($\neg$ e(e)))
\end{lstlisting}
Expression ($\star$) is inconsistent for the model theory given in the next section as it must be in every well-specified model theory because it is self-contradictory. Thus, expression ($\star$) is, like the above specification of the  Barber Paradox a mere inconsistency. 

While it is paradoxical to think of concepts that cannot exist, inconsistent expressions are no reasons to reject the language in which they are expressed. After all, nobody considers the language of propositional logic as paradoxical, notwithstanding the fact that it can express the formula \lstinline{(p $\land$ ($\neg$ p))} which is inconsistent for requiring a proposition \lstinline{p} to be both true and false. 

The rejection of logics in which Russell's paradox can be expressed stems from the conception that every expression must define a set. While it is understandable that Russell, Frege and their contemporaries shared this conception, this conception can be given up. Giving up this conception provides for a simple logic perfectly formalising Prolog-style meta-programming. Giving up this conception allows for impredicative atoms that are interpreted as collection like the collection of beliefs mentioned above in Section \ref{sec:Predicativity-Impredicativity}. 

\section{Variant Expressions and Expression Rectification}\label{sec:Variant-Rectification} 

In the next section, a model theory is given for Reflective Predicate Logic such that two syntactically distinct atomic expressions that are variants of each other like
\begin{lstlisting}
    believes(ann, ($\forall$ X believes(bill, X)))
    believes(ann, ($\forall$ Y believes(bill, Y)))
\end{lstlisting} 
(with the intended meaning that Ann believes that Bill believes everything) are identically interpreted. As already mentioned, this is not the case with the model theory of Ambivalent Logic \cite{vademecum-1,vademecum-2}. 

Two first-order logic atoms or terms $E_1$ and $E_2$ are variants of each other if there is a one-to-one mapping $\sigma$ of the variables occurring in $E_1$ into the variables occurring in $E_2$ such that applying $\sigma$ to $E_1$ yields $E_2$, noted $E_1 \sigma = E_2$. Thus, the first-order formulas
\begin{lstlisting}
   p(X, Y) 
   p(Y, Z)
\end{lstlisting} 
(in which \lstinline{X}, \lstinline{Y} and \lstinline{Z} are first-order variables) are variants of each other but the first-order formulas
\begin{lstlisting}
   p(X, Y)  
   p(Z, Z)
\end{lstlisting}  
(in which \lstinline{X}, \lstinline{Y} and \lstinline{Z} are first-order variables) are not. 

Variance is more complex to formalise for first-order logic formulas and Reflective Predicate Logic expressions because of the overriding (or variable shadowing, or shadowing, for short) that might take place with quantification. While the first-order formulas 
\begin{lstlisting}
    (p(X) $\land$ q(Y)) 
    (p(X) $\land $ q(X))
\end{lstlisting}  
(in which \lstinline{X} and \lstinline{Y} are first-order variables) are not variants of each other, the first-order formulas 
\begin{lstlisting}
    ($\forall$ X (p(X) $\land$ $\exists$ Y q(Y))) 
    ($\forall$ X (p(X) $\land$ $\exists$ X q(X)))
\end{lstlisting}    
are variants of each other because, in the second formula, the second quantification of the variable \lstinline{X} overrides the first. 

Variant expressions can easily be defined by relying on a rectification. Rectifying an expression consists in renaming its variables from a predefined pool of ``fresh'' variables, that is, variables not occurring in the expressions considered, in such a manner that the variables of two distinct quantifications are distinct. Thus, if the ``fresh'' variables considered are $v_1, v_2, \ldots$, then rectifying the expression
\begin{lstlisting}
    ($\forall$ X (p(X) $\land$ $\exists$ X q(X)))
\end{lstlisting} 
might result in    
\begin{lstlisting}
    ($\forall$ $v_1$ (p($v_1$) $\land$ $\exists$ $v_2$ q($v_2$)))
\end{lstlisting} 
and 
rectifying the expression
\begin{lstlisting}
    (($\forall$ X p(X)) $\land$ ($\exists$ X q(X)))
\end{lstlisting} 
might result in       
\begin{lstlisting}
    (($\forall$ $v_3$ p($v_3$)) $\land$ ($\exists$ $v_1$ q($v_1$)))
\end{lstlisting} 

It is assumed in the following that there is a denumerable supply $v_1, v_2, \ldots, v_i,$ $\ldots$ of variables such that each $v_i$ is distinct from every logical and every non-logical symbol of the Reflective Predicate Logic language considered. (An infinite supply of variables is necessary to ensure that proofs are not bounded in length. This infinity does not threaten computability because the variables needed in a proof can be created on demand while computing proofs.)

A rectification is performed by a variable renaming which is conveniently specified as a predicate {\it rect} recursively defined on an expression's structure. 
It implements a left-to-right outside-in traversal of expressions of all kinds except quantified expressions that are traversed inside-out so as to reflect the quantified variables' scopes: Note, in the last case of the algorithm, the recursive call $\textit{rect}(E_2, i, R_2, j)$ instead of $\textit{rect}(E_1, i, R_1, j)$. 
Logic programming pseudo-code is used in the following definition because it expresses the sideway passing of variable indices between recursive calls in a more readable manner than functional pseudo-code.

\begin{definition}[Rectification]\label{def:rectification-algorithm}
A rectified $R$ of an expression $E$ 
is specified by the predicate
  $\textit{rect}(E, i, R, j)$ where: 
\begin{itemize}
   \item $i \geq 1$, the ``initial variable index'', denotes the first variable $v_i$ that might be used in rectifying $E$
   \item $j = i$ if $E$ and its rectified contain no variables
   \item $j > i$, the ``final variable index'', is $k+1$ if $k$ is the highest index of a variable occurring in the rectified of  $E$ 
\end{itemize}
The predicate $\textit{rect}$ is recursively defined on an expression's structure:
\begin{itemize}
   \item if $E$ is a symbol: \\
     $\textit{rect}(E, i, E, i)$
   \item if $E = E_1(E_2, \ldots, E_n)$ with $n \geq 2$: \\
     $\textit{rect}(E, i, R_1(R_2, \ldots, R_n), i_n)$ where the $R_k$ and $i_n$ are defined by \\  
     $\textit{rect}(E_1, i, R_1, i_1)$, 
      $\textit{rect}(E_2, i_1, R_2, i_2)$, $\ldots$, and 
      $\textit{rect}(E_n, i_{n-1}, R_n, i_n)$
   \item if $E = (\neg E_1)$: \\
     $\textit{rect}(E, i, (\neg R_1), j)$ is defined by $\textit{rect}(E_1, i, R_1, j)$
   \item if $E = (E_1 \theta E_2)$ with $\theta \in \{\land, \lor, \Rightarrow\}$: \\
    $\textit{rect}(E, i, (R_1 \theta R_2), i_2)$ where $R_1$, $R_2$ and $i_2$ are defined by \\
       $\textit{rect}(E_1, i, R_1, i_1)$ and 
       $\textit{rect}(E_2, i_1, R_2, i_2)$
   \item if $E = (\theta~ E_1~ E_2)$ with $\theta \in \{\forall, \exists\}$: \\
    $\textit{rect}(E, i, (\theta~ v_j~ R), k)$ where $R$, $j$ and $k$ are defined by \\
    $\textit{rect}(E_2, i, R_2, j)$, 
       $k = j + 1$, and 
       $R$ is obtained from $R_2$ by simultaneously replacing all occurrences of $E_1$ in $R_2$ by $v_j$. 
\end{itemize}

A rectified 
of a finite set $\{E_1, \ldots, E_n\}$ of expressions is the set  $\{R_1, \ldots, R_n\}$ where $(R_1 \wedge (\ldots \wedge R_n)\ldots)$ is the rectified  
of the expression$(E_1 \wedge (\ldots \wedge E_n)\ldots)$.

An expression (set of expressions, respectively) is said to be rectified if it is a rectified  of some expression (set of expressions, respectively). 
\end{definition}

Because of the inside-out traversal of quantified expressions, the predicate {\it rect} is not tail recursive. An efficient, tail recursive, implementation of {\it rect} would require an accumulator for storing embedding quantified expressions the variable renaming of which is delayed. 

The algorithm specified in Definition \ref{def:rectification-algorithm} terminates because every recursive call refers to a strict sub-expression. $\textit{rect}(E, i, R, j)$ is functional in the sense that there is exactly one pair $(R, j)$ for each pair $(E, i)$ because the cases of the above definition are mutually exclusive. The call pattern of $\textit{rect}$ is $\textit{rect}(+E, +i, ?R, ?n)$ meaning that in a call to {\it rect} $E$ and $i$ must be specified and that each of $R$ and $n$ can, but do not have to, be specified. 
Therefore, it is also possible to define rectification functionally with a binary mapping $\textit{rect}: \textit{expr} \times \mathbb{N} \rightarrow \textit{expr} \times \mathbb{N}$ where $\textit{expr}$ is the set of all expressions of a Reflective Predicate Logic language. 
No variables are overridden in the rectified of an expression because of the sideway passing of variable indices between recursive calls: Each recursive call to {\it rect} uses as initial variable index the final variable index of the previous recursive call to {\it rect}. Let $R_i$ denote the rectified of an expression $E$ specified  by $\textit{rect}(E, i, R_i, n)$. For all $k \geq 1$ $R_{i+k}$ can be obtained from $R_i$ by replacing in $R_i$ every variable $v_j$  by $v_{j+k}$. 

In general, a finite set of expressions has more than one rectified, each resulting from an ordering of the set's expressions. Note that the expressions in a rectified set of expressions are standardised apart, that is, two distinct expressions have no variables in common. 

In contrast to first-order formulas' rectification, the algorithm of Definition \ref{def:rectification-algorithm} considers the variables that might occur in the constructors of Reflective Predicate Logic atoms. 
The Reflective Predicate Logic expression 
\begin{lstlisting}
    (($\forall$ X p(X)) $\land$ ($\exists$ X q(X)))
\end{lstlisting} 
is for example rectified by the algorithm of Definition \ref{def:rectification-algorithm} into: 
\begin{lstlisting}
    (($\forall$ $v_2$ p($v_2$)) $\land$ ($\exists$ $v_1$ q($v_1$)))
\end{lstlisting} 
while the Reflective Predicate Logic atom
\begin{lstlisting}
    ($\forall$ T (trust(T) $\Rightarrow$ T))(ann, bill)
\end{lstlisting}
is rectified by the same algorithm into: 
\begin{lstlisting}
    ($\forall~ v_1$ (trust($v_1$) $\Rightarrow v_1$))(ann, bill)
\end{lstlisting}

\begin{definition}[Variant Expressions]\label{def:variant-expressions}
Two expressions $E_1$ and $E_2$ are variants of each other, noted $E_1 {\sim} E_2$, if their rectified after Definition \ref{def:rectification-algorithm} and computed with the same initial variable index are identical. 
\end{definition}
The relation $\sim$ on the expressions of a Reflective Predicate Logic language is an equivalence relation. 

\section{A Herbrand-Style Model Theory for Reflective Predicate Logic}\label{sec:Model-Theory}

Atoms in Reflective Predicate Logic (that is, expressions the outermost constructors of which are neither connectives nor quantifiers) like 
\begin{lstlisting}
   likes(ann, likes(bill, ann)) 
   likes(ann, ($\forall$ x likes(bill, x)))
   (likes $\wedge$ trusts)(ann, bill)
   ($\forall$ T trust(T) $\Rightarrow$ T)(ann, bill)
\end{lstlisting}
(meaning that Ann likes that Bill likes her, that Ann likes that Bill likes everyone and everything, that Ann likes and trusts Bill, and that Ann trusts Bill in all specified forms of trust) differ from atoms in a first-order logic language in three respects: 
\begin{enumerate}
   \item First-order atoms may be open or closed, whereas all Reflective Predicate Logic expressions, including all atoms, are closed expressions thanks to the paradigm ``quantification makes variables''.
   \item First-order logic atoms cannot have anything but terms as arguments, whereas atoms in Reflective Predicate Logic may have as arguments non-atomic (or logical) expressions such as \lstinline{($\forall$ x likes(bill, x))} that amount to first-order formulas, not terms. 
   \item In first-order logic the outermost constructor of an atom is a symbol such as \lstinline{likes}, whereas in Reflective Predicate Logic it can be any expression
   such as  \lstinline{likes} and \\
   \lstinline{($\forall$ T trust(T) $\Rightarrow$ T)}.
\end{enumerate}

How the model theory should treat atoms that contain non-atomic, or logical, expressions can be seen on the following example the meaning of which is that calling someone ``A and B'' implies calling him ``A'' and calling him ``B'', that claiming not to call someone ``A'' is in fact calling him ``A'' and that calling someone ``fat'' is offending. 
\begin{lstlisting}
   ($\forall$ x ($\forall$ y ($\forall$ z 
      (says(x, (y $\land$ z)) $\Rightarrow$ (says(x, y) $\land$ says(x, z))))))
   ($\forall$ x ($\forall$ y (says(x, ($\neg$ says(x, y))) $\Rightarrow$ says(x, y))))
   ($\forall$ x ($\forall$ y (says(x, is(y, fat)) $\Rightarrow$ offends(x, y)))) 
\end{lstlisting}
An interpretation satisfying the three above expressions as well as the additional atom 
\begin{lstlisting}
   says(donald, ($\neg$says(donald, is(kim, (short$\land$fat))))
\end{lstlisting}
should also satisfy
\begin{lstlisting}
   offends(donald, kim)
\end{lstlisting}
regardless of whether none, only one, or both of 
\begin{lstlisting}
   is(kim, (short $\land$ fat))
   is(kim, fat)
\end{lstlisting}
are satisfied in that interpretation \cite{trump-fat-and-short}. 
This requirement is essential among others for static program analyses (like static type checking) to be expressible as meta-programs. Indeed, a static program analysis is independent of the analysed programs' run time behaviours, that is, a static analysis of a logic program is independent of which program parts evaluate to true. 
The model theory specified below is tuned to ensure this requirement.  

In contrast to first-order logic ground atoms, Reflective Predicate Logic atoms can have variants, like 
\begin{lstlisting}
   believes(ann, ($\forall$ y     (believes(ann, y   ) 
                             $\Rightarrow$ believes(bill, y     )))) 
   believes(ann, ($\forall$ t(a) (believes(ann, t(a)) 
                             $\Rightarrow$ believes(bill, t(a)))))
\end{lstlisting}
that should be given the same meaning even though they syntactically differ from each other. Thus, the Herbrand  base of a Reflective Predicate Logic language must be defined as the set of equivalence classes of the language's atoms with respect to the variant relation $\sim$. 

Since atoms like 
\begin{lstlisting}
   believes(ann, ($\forall$x (believes(ann,x)$\Rightarrow$believes(bill,x))))
\end{lstlisting}
(with the intended meaning that Ann believes that Bill believes all that she herself believes) of a Reflective Predicate Logic language correspond to both ground atoms and ground terms of first-order logic languages, the set of all expressions of a Reflective Predicate Logic language corresponds to both the Herbrand universe \cite{fixpoint-semantics-1,Chang-Lee} and the Herbrand base \cite{fixpoint-semantics-1,Chang-Lee} of a first-order logic language. 

\begin{definition}[Herbrand Universe]\label{def:universe}
Let $\cal A$ be the set of atoms of a Reflective Predicate Logic language $\cal  L$ and $\sim$ the variant relation of $\cal L$. The Herbrand universe of $\cal  L$ is $\cal A /\sim$, that is, the set of equivalence classes of $\sim$. 
\end{definition}

As the following example illustrates, standardisation-apart is needed in proving so as to properly reflect variable scopes. In this example, upper case characters denote variables. From the clauses 
\begin{lstlisting}
    $[\neg p(X), \neg q(Y), r(X, Y)]$ 
    $[p(Z)]$
    $[q(Z)]$ 
\end{lstlisting}  
the clause 
\begin{lstlisting}
   $[r(Z_1, Z_2)]$
\end{lstlisting}  
can be derived by resolution.  
If, however, no standardisation-apart of the set of clauses was performed during resolution, then the more specific clause ``$[r(Z, Z)]$'' (or a variant of that clause) would be wrongly derived instead of ``$[r(Z_1, Z_2)]$'' (or a variant of that clause). 

Under the paradigm ``quantification makes variables'' something similar might happen while instantiating variables. Consider once again the Reflective Predicate Logic expression 
\begin{lstlisting}
    $ (\dagger)~ (\forall$ bill ($\exists$ ann likes(ann, bill)))
\end{lstlisting}  
(meaning that everyone is liked by someone) in which the symbols \lstinline{ann} and \lstinline{bill} serve as variables. Prematurely instantiating \lstinline{bill} with \lstinline{ann} yields 
\begin{lstlisting}
    ($\exists$ ann likes(ann, ann))
\end{lstlisting}  
(meaning that someone likes herself) which is not a logical consequence of $(\dagger)$. Such incorrect instantiations are avoided by rectifying the expressions under consideration using the infinite supply of variables $v_1, v_2, \ldots$ that has been assumed in the previous section \ref{sec:Variant-Rectification}. Indeed, rectified expressions do not contain non-variable symbols serving as variables. Since each variable $v_i$ is distinct from every non-logical symbol as well as from every logical symbol of the Reflective Predicate Logic language considered, incorrect instantiations like in the former example are impossible. 

\begin{definition}[Notations]
If $A$ is an atom of a Reflective Predicate Logic language $\cal L$, then class$(A)$ denotes the variant class of $A$, that is, the equivalence class of $A$ in the Herbrand universe $\cal A /\sim$ of $\cal  L$. 

If $R$ is a rectified expression, if $v_i$ is a variable and if $A$ is an atom, then $R[A/v_i]$ denotes the expression obtained from standardised-apart rectified variants $R^v$ and $A^v$ of $R$ and $A$ respectively by simultaneously replacing in $R^v$ all occurrences of  $v_i$ by $A^v$.
\end{definition}

In defining the notation $R[A/v_i]$, there is no need for caring about overridden variables because that notation will only apply to rectified expressions $R$ and because, as observed in the previous section \ref{sec:Variant-Rectification},  in rectified expressions no variables are overridden. Since $R^v$ and $A^v$ are rectified and have no variables in common (they are standardised apart), $R[A/v_i]$ is rectified. 

\begin{definition}[Interpretations and Models]\label{def:interpretation}
A Herbrand interpretation $I(S)$ of a Reflective Predicate Logic language $\cal  L$ is specified as a subset $S$ of the universe $\cal A /\sim$ of $\cal L$.

An expression $E$ is satisfied in a Herbrand interpretation $I(S)$ of $\cal  L$, denoted $I(S) \models E$, if a rectified $R$ of $E$ is satisfied in $I(S)$, denoted $I(S) \models R$, in the following sense, where:
\begin{itemize}
   \item $R, R_1$, and $R_2$ denote \emph{rectified} expressions.
   \item $A$ denotes a \emph{rectified} atom. 
\end{itemize}

\begin{tabular}{ l l l }
   $I(S) \models A$                				& iff & class$(A) \in S$ 	                                                     	\\
   $I(S) \models (\neg R)$            				& iff & $I(S) \not\models R$						\\
   $I(S) \models (R_1 \land R_2)$      			& iff & $I(S) \models R_1$ and $I(S) \models R_2$ 		\\
   $I(S) \models (R_1 \lor R_2)$     			& iff & $I(S) \models R_1$ or $I(S) \models R_2$   		\\
   $I(S) \models (R_1 \Rightarrow R_2)$~     		& iff & if $I(S) \models R_1$, then $I(S) \models R_2$ 	\\
   $I(S) \models (\exists v_i~ R)$			 	& iff & $I(S) \models R[A/v_i]$ for some $A$    			\\	
   $I(S) \models (\forall v_i~ R)$   				& iff & $I(S) \models R[A/v_i]$ for all $A$             		\\
\end{tabular}

\vspace{0.5em}

A set $T$ of expressions is satisfied in $I(S)$, denoted $I(S) \models T$, if every expression in $T$ is satisfied in  $I(S)$. 

An interpretation is called a model of an expression $E$ (a set of expressions $P$, respectively) if it satisfies $E$ (every expression in $P$, respectively).
\end{definition}

Satisfaction of logical expressions (that is, expressions the outermost symbols of which are logical symbols ($\neg$, $\wedge$, $\vee$, $\Rightarrow$, $\forall$, $\exists$)) is defined like in first-order logic. Satisfaction of atoms (that is, expressions the outermost expressions of which are non-logical symbols) is not defined like in first-order logic: It is based on variance instead of syntactical identity. However, if an atom does not contain variables, then it is the single element of its variant class and, as a consequence, its satisfaction is defined like in first-order logic. 

Applied to first-order logic expressions, Definitions \ref{def:universe} and \ref{def:interpretation} amount to the definitions of Herbrand universes, interpretations and models of first-order logic. Indeed, a ground atom $A$ of a first-order logic language is the only element of its equivalence class for the variant relation. Thus, Definition \ref{def:interpretation} is a conservative extension of first-order logic's notions of Herbrand interpretations and models. This observation is formally developed in the next section \ref{sec:Conservative-Extension}.

An interpretation as defined above can be seen as a set $S$ of atoms. An atom is satisfied in the interpretation specified by $S$ if and only if it is a variant of an element of $S$. 

Consider the following set $P_1$ of expressions, a simple meta-program (with explicit quantifications) on the beliefs of Ann and Bill:  
\begin{lstlisting}
 believes(ann, itRains) 
 believes(ann, (itRains $\Rightarrow$ itIsWet)) 
 believes(ann,($\forall$x (believes(ann,x) $\Rightarrow$ believes(bill,x)))) 
 ($\forall$ x (believes(ann, x) $\Rightarrow$ believes(bill, x)))
\end{lstlisting}
The following set $S_1$ of atoms specifies a model of $P_1$ (consisting of the atoms' equivalence classes for $\sim$): 
\begin{lstlisting}
 believes(ann, itRains) 
 believes(ann, (itRains $\Rightarrow$ itIsWet)) 
 believes(ann,($\forall$y (believes(ann,y) $\Rightarrow$ believes(bill,y)))) 
 believes(bill,itRains) 
 believes(bill,(itRains $\Rightarrow$ itIsWet)) 
 believes(bill,($\forall$z (believes(ann,z) $\Rightarrow$ believes(bill,z)))) 
\end{lstlisting}

Consider the following set $P_2$ of expressions in which
\lstinline{($\forall$ T (trust(T) $\Rightarrow$ T))} is an atom constructor:  
\begin{lstlisting}
 trust(t1)
 trust(t2)
 t1(ann, bill)
 t2(ann, bill)
 ($\forall$ X ($\forall$ Y 
     ( ($\forall$ T (trust(T) $\Rightarrow$ T(X, Y))) $\Rightarrow$  
       ($\forall$ T (trust(T) $\Rightarrow$ T))(X, Y) 
     ) 
   )
 )
\end{lstlisting}
The following set $S_2$ of atoms, in which \lstinline{($\forall$ X (trust(X) $\Rightarrow$ X))} is an atom constructor, specifies a model of $P_2$:
\begin{lstlisting}
 trust(t1)
 trust(t2)
 t1(ann, bill)
 t2(ann, bill) 
 ($\forall$ X (trust(X) $\Rightarrow$ X))(ann, bill)
\end{lstlisting}
A proof theory convenient for expressions like the last of $P_2$ is out of the scope of this article. 

The definition of an interpretation given in the articles \cite{vademecum-1,vademecum-2} is more stringent than Definition \ref{def:interpretation}: Instead of relying on the variant relationship, it requires syntactical identity. As a consequence, the set $S_1$ of atoms given above does not specify a model in the sense of \cite{vademecum-1,vademecum-2} of the set $P_1$ of expressions given above. This is undesirable because meta-programming requires to interpret identically expressions like the following that are variants of each other: 
\begin{lstlisting}
 believes(ann,($\forall$x (believes(ann,x) $\Rightarrow$ believes(bill,x))))
 believes(ann,($\forall$y (believes(ann,y) $\Rightarrow$ believes(bill,y)))) 
\end{lstlisting}

Even though a Reflective Predicate Logic language gives rise to inconsistent expressions (like the definition ($\star$) of Russell's paradoxical set given in Section \ref{sec:Paradoxes}), the model theory given above is adequate for the reasons  mentioned at the end of Section \ref{sec:Predicativity-Impredicativity}. Indeed, it refers to an inductively defined universe, the (standard) set of all expressions, and impredicative atoms like 
\begin{lstlisting}
 believes(ann,($\forall$x (believes(ann,x)$\Rightarrow$believes(bill,x))))
 belief(belief(itRains))
\end{lstlisting}
perfectly characterise elements of that universe even though they cannot be interpreted as standard sets. 

\noindent 
\section{Symbol Overloading in Classical Predicate Logic Langua\-ges}
\label{sec:Symbol-Overloading}

The clause
\begin{lstlisting}
   student(X) :- enrolled(student(X), _)
\end{lstlisting}  
can be seen as a first-order logic clause even though under this view the symbol \lstinline{student} occurs in that clause both as a predicate symbol (left) and as a function symbol (right). 
Indeed, even though first-order logic languages are conventionally defined such that a same (predicate or function) symbol cannot be used with different arities and a same symbol cannot be used both as a function and as a predicate symbol, these conventions are not necessary for two reasons: First, arities can be seen as part of a symbol which allows to disambiguate occurrences of a same symbol with different arities; second the syntactic context in a formula allows to disambiguate occurrences of a same symbol denoting a function symbol and a predicate symbol.

The following example illustrates this observation. Consider 
\begin{itemize}
   \item a function symbol \lstinline{e} of artity 1 meant to express a function ``twice'', that is, \lstinline{e(1)} stands for $2 \times1$. 
   \item a function symbol \lstinline{e} of artity 2 meant to express the addition of integers, that is, \lstinline{e(1,1)} stands for $1 + 1$. 
   \item a predicate symbol \lstinline{e} of arity 1 meant to express ``even'', that is, \lstinline{e(1)} stands for  \lstinline{isEven(1)}.
   \item a predicate symbol \lstinline{e} of arity 2 meant to express equality, that is, \lstinline{e(1,1)} stands for $1 = 1$.
\end{itemize}

If \lstinline{e(1)} is a first-order  term, then it can only stand for $2 \times 1$. 
If \lstinline{e(1,1)} is a first-order term, then it can only stand for $1 + 1$. 
If \lstinline{e(e(1,1),e(1))} is a first-order  formula, then the outer occurrence of \lstinline{e} must be a predicate symbol, the inner occurrences of \lstinline{e} must be function symbols and the formula can only stand for  $1 + 1 = 2 \times 1$. 
Similarly, if \lstinline{e(e(e(1,1)))} is a first-order formula, then it can only stand for  \lstinline{isEven(2 $\times$ (1 + 1))}. 

The above observation is common knowledge in logic and in programming. 
In logic, Church's re-formulation of the Simple Type Theory \cite{Chwistek-Simple-Type-Theory-Polish,Ramsey-Simple-Type-Theory}, the Simply Typed Lambda Calculus  \cite{church-lambda-calculus,typed-lambda-calculus}, exploits it. The above observation has been made for classical predicate logic in the article \cite{hilog}. In that article, a predicate logic language overloading symbols in the aforementioned sense is called ``contextual.'' 
In programming, the use a same symbol with different meanings in different contexts is called overloading. 
In functional and Prolog programs, 
a same symbol is commonly used with different arities. In functional programs type selectors are commonly named like the corresponding type constructors and in Haskell compound types are named using their own type constructor (\lstinline{[Int]} for example denotes the type of the lists of integers). 

Thus, interpreting in first-order logic a clause like
\begin{lstlisting}
   student(X) :- enrolled(student(X), _)
\end{lstlisting}
in which some symbols are interpreted both as function and predicate symbols requires no further formalisation and the  aforementioned symbol overloading in first-order languages accounts for the aforementioned Prolog facts: 
\begin{lstlisting}
   call(twice, 1, 2)

   forall(enrolled(student(S), mathematics), 
                  attends(S, logic)
   )

   forall(enrolled(student(S), P), 
                  forall(syllabus(P, C), attends(S, C)
                  )
   )
\end{lstlisting}

However, the  aforementioned symbol overloading in first-order languages accounts neither for the representation in classical predicate logic of \lstinline{call(P, X, Y)}, that is, \lstinline{P(X, Y)}, nor for the following Prolog meta-program that defines the \lstinline{forall} predicate: 
\begin{lstlisting}
   forall(R, F) :- not (R, not F)
\end{lstlisting}
Indeed, the aforementioned symbol overloading in first-order languages does not lift the strong typing of classical predicate logic discussed in the next section \ref{sec:Related-Work}:  In classical predicate logic, a variable cannot range over both terms and formulas. The following example is a another case of confounding of object and meta-variables that cannot be accounted for in classical predicate logic by the aforementioned symbol overloading: 
\begin{lstlisting}
   r(X) :- X(X)
\end{lstlisting}
For the same reason, Russell's Paradox ($\star$) is not expressible in a first-order logic language sharing symbols.

\section{Reflective Predicate Logic is a Conservative Extension of First-Order Logic}\label{sec:Conservative-Extension}

Reflective Predicate Logic's syntax differs from that of first-order logic in having only one category of expressions, whereas first-order logic distinguishes between terms and formulas. First-order logic terms can be expressed in Reflective Predicate Logic as follows: 

\begin{definition}[First-Order Terms]\label{def:terms}
Let $\cal L$ be a Reflective Predicate Logic language the set of non-logical symbols of which is $S$. 
A first-order term fragment ${\cal F}_{t}$ of $\cal L$ is specified by: 
\begin{itemize}
   \item A set $V \subseteq S$ of variables and a set $T \subseteq S$ of term symbols such that $V \cap T = \emptyset$
   \item The assignment to each element of $T$ of at least one arity (that is, non-negative integer)
\end{itemize}
The terms of ${\cal F}_{t}$ are inductively defined as follows: 
\begin{itemize}
   \item A variable is a term. 
   \item A term symbol with arity $0$ is a term. 
   \item If $f$ is a  term symbol with arity $n \geq 1$ and $t_1, \ldots, t_n$ are terms, then $f(t_1, \ldots, t_n)$ is a term. 
\end{itemize}
\end{definition}

Term symbols with arity $0$ are commonly called ``constants'', term symbols with arity $n \geq 1$ ``function symbols''. It is commonly required in mathematical logic that each term symbol has exactly one arity. This requirement is not necessary as it is recalled above in Section \ref{sec:Symbol-Overloading}. 

Neglecting that Reflective Predicate Logic requires parentheses around quantified and negated expressions that are superfluous in first-order logic, first-order logic formulas can be expressed in Reflective Predicate Logic as follows:

\begin{definition}[First-Order Formulas]\label{def:formulas}
Let $\cal L$ be a Reflective Predicate Logic language the set of non-logical symbols of which is $S$. 
A first-order fragment $\cal F$ of $\cal L$ is specified by a first-order term fragment of $\cal L$ with set of variables $V$ and set of term symbols $T$ and by
\begin{itemize}
   \item A set $P \subseteq S$ of predicate symbols such that $V \cap P = \emptyset$
   \item The assignment to each element of $P$ of at least one arity (that is, non-negative integer)
\end{itemize}
The first-order formulas of $\cal F$ are inductively defined as follows: 
\begin{itemize}
   \item A predicate symbol with arity $0$ is a formula. 
   \item If $p$ is a predicate symbol with arity $n \geq 1 $ and $t_1, \ldots, t_n$ are terms, then $p(t_1, \ldots, t_n)$ is a formula. 
     \item  If $F$ is a formula, then $(\neg F)$ is a formula. 
     \item If $F_1$ and $F_2$ are formulas, then $(F_1 \land F_2)$, $(F_1 \lor F_2)$, $(F_1 \Rightarrow F_2)$ are formulas. 
     \item If $x$ a variable and $F$ is a formula, then $(\forall x~  F)$ and $(\exists x~  F)$ are formulas. 
\end{itemize}
A formula $F$ is open if a variable $x$ occurring in a subexpression of $F$ is not in the scope of a quantification of the form $(\forall x E)$ or $(\exists x E)$. A formula is closed, or a sentence, if it is not open. 
\end{definition}

 It is commonly required in mathematical logic that each predicate symbol has exactly one arity and that $P \cap T= \emptyset$. As it is recalled above in Section \ref{sec:Symbol-Overloading}, these requirements are not necessary. 

A closed formula which is an atomic expression, or atom, is commonly called a ``ground atom''. 

Definitions \ref{def:terms} and \ref{def:formulas} use and constrain rules of Definition \ref{def:expressions}. As a consequence, terms and formulas of a first-order logic fragment of a Reflective Predicate Logic language are expressions of that language. 

\begin{definition}[First-Order Herbrand Interpretations]\label{def:FO-interpretations}
Let $\cal F$ be a first-order fragment of a Reflective Predicate Logic language $\cal L$. 
A first-order Herbrand interpretation of $\cal F$ is specified as a set of ground atoms of $\cal F$. 
Satisfiability in a first-order Herbrand interpretation of $\cal F$ is defined as in Definition \ref{def:interpretation}. 
If $I$ is a Herbrand interpretation of $\cal L$, the restriction of $I$ to the first-order fragment $\cal F$ of $\cal L$, noted $I_{\cal _F}$, is the subset of $I$ consisting of the variant classes of atoms of $\cal F$. 
\end{definition}

Observe that if $I$ is a Herbrand interpretation of a Reflective Predicate Logic language $\cal L$ and if $\cal F$ is a first-order fragment of $\cal L$, then the restriction of $I$ to $\cal F$, $I_{\cal _F}$, is a Herbrand interpretation in the sense of Definition \ref{def:interpretation}. 
Herbrand interpretations of first-order logic languages are usually specified as sets of ground atoms, not sets of ground atoms' variant classes. However, since a ground atom is the single element of its variant class, Herbrand interpretations of first-order logic languages can be seen as sets of variant classes. 

\begin{proposition}[Conservative Extension]\label{prop:Conservative-Extension}
Consider 
\begin{itemize}
   \item $\cal F$ a first-order language with set of variables $V$, set of term symbols $T$ and set of predicate symbols $P$
   \item ${\cal F}_{RPL}$ the Reflective Predicate Logic language with set of non-logical symbols $V \cup T \cup P$
   \item $I$ a Herbrand interpretation of $\cal F$
   \item $F$ a formula of $\cal F$
 \end{itemize}
Reflective Predicate Logic is a conservative extension of first-order logic, that is: 
\begin{itemize}
   \item $F$ is an expression of ${\cal F}_{RPL}$.
   \item $I \models_{FOL} F$ if and only if for all Herbrand interpretations $J$ of ${\cal F}_{RPL}$ such that $J_{\cal F} = I$, $J \models_{RPL} F$.
\end{itemize}
where $\models_{FOL}$ and $\models_{RPL}$ denote satisfiability in first-order and Reflective Predicate Logic interpretations respectively. 
\end{proposition}

\paragraph{Proof.} 
The first point has been already observed above. The second point follows from the fact that satisfiability of a formula (or expression) in an interpretation is defined recursively on the formula's (or expression's) structure. The satisfiability of $F$ in $J$ therefore depends only on expressions built from $\cal F$'s vocabulary, that is, depends only on the subset $J_{\cal F} = I$ of $J$.

\section{Conclusion}\label{sec:Conclusion}

This article has given Prolog-style meta-programming, which is characterised by a confounding of terms and formulas and of object and meta-variables, a simple formalisation, arguably the simplest, the most complete and the closest to the programming practice so far proposed. This formalisation consists in a systematisation of the syntax of Frege's logic, the precursor of classical predicate logic, and in a generalisation of Herbrand model theory. The resulting logic, Reflective Predicate Logic, has been shown to be a conservative extension of first-order logic. 

The aforementioned syntax systematisation is simple: It consists in drawing the consequences of replacing expressions by their definitions and in an unconventional representation of variables easing meta-programming. This syntax systematisation had been initiated in the article \cite{hilog} and completed in the articles \cite{vademecum-1,vademecum-2}. 

The aforementioned generalisation of Herbrand model theory is simple, too. It consists in considering reflective atoms, that is, atoms that might contain connectives, quantifiers, and variables, instead of ground atoms. This, too, had been initiated in the article \cite{hilog} and furthered, though in an unsatisfying manner, in the articles  \cite{vademecum-1,vademecum-2}. A simple change, the identical interpretation of variant expressions, was enough to yield a satisfying Herbrand-style model theory.  

The resulting logic, Reflective Predicate Logic, is simple. It differs from first-order logic in only three rather simple aspects.  The syntax of Reflective Predicate Logic renounces type theory, which makes it simpler than the syntax of first-order logic. The syntax of Reflective Predicate Logic has an unconventional representation of variables. The model theory of Reflective Predicate Logic is based on a generalisation of Herbrand model theory which gives rise to specify collections in Reflective Predicate Logic, generalisations of sets which perfectly interpret reflective expressions. 

In spite of its simplicity, Reflective Predicate Logic is significant to logic programming, know\-ledge representation and mathematical logic. Reflective Predicate Logic is significant to logic programming because it accommodates Prolog-style meta-progra\-mming without the restrictions required by the formalisations of meta-programming in higher-order logic, without the restrictions and without the encodings (or naming relations) required by the formalisations of meta-programming in first-order logic, and without the inadequacies of HiLog and Ambivalent Logic (discussed in Section \ref{sec:Related-Work}). Reflective Predicate Logic is also significant to logic programming because it provides a justification to the confounding of object and meta-variables, the so-called ``non-ground representations'' (mentioned in Section \ref{sec:Related-Work}) of efficient deduction systems. Reflective Predicate Logic is significant to knowledge representation because it is natively reflective thus considerably simpler than the standard approaches to reflection that are based on reification. Reflective Predicate Logic is significant to mathematical logic because it is an alternative to type theory and a rehabilitation of Frege's logic. The simplicity of Reflective Predicate Logic contributes to its significance. Indeed, in science, simplicity is not a drawback but instead an advantage. 

A further contribution of this article is its handling of Russell's Paradox of self-reflectivity by proposing a logic in which the paradox is expressible. The widespread common wisdom is instead that a well-defined logic should preclude paradoxes. A paradox is a counter-intuitive inconsistency. A thesis of this article is that there is nothing problematic with a logic in which inconsistencies, be they intuitive or counter-intuitive, can be expressed. 

This article is a first step. Further work should be devoted to:
\begin{itemize}
   \item generalising the model theory of this article to universes of all kinds, possibly including universes that are collections in the sense of Section \ref{sec:Predicativity-Impredicativity} instead of sets, 
   \item giving Reflective Predicate Logic a unification and a resolution calculus,
   \item specifying a logic programming syntax, preferably with a type system, based on Reflective Predicate Logic,
   \item investigating how structuring constructs such as modules and embedded implications \cite{theory-of-modules,modules-in-lp,hilog,Giordano-1994-Embedded-Implications,Giordano-1994-Structured-Prolog,Haemmerle-2006-Modules-for-Prolog-Revisited} can be formalised in Reflective Predicate Logic, 
   \item investigating how paradoxes of self-reflection, especially those used in proving G\"{o}del's incompleteness theorems, can be expressed in Reflective Predicate Logic. 
\end{itemize}

\section*{Acknowledgments}

The author is thankful to Norbert Eisinger, Bob Kowalski and Antonius Weinzierl for fruitful discussions on the subject of this article. The author acknowledges useful hints from the journal area editor Michael J.\ Maher, from the anonymous reviewers
and from the audience of the 21st International Conference on Applications of Declarative Programming and Knowledge Management (INAP 2017) during which part of the work reported about in this article has been presented in a talk without associated publication. The author is thankful to Norbert Eisinger and Elke Kroi{\ss} for their help in correcting typos and stylistic lapses in drafts of this article. 

\section*{Appendix: A Brief Introduction into Frege's logic}

This brief introduction into Frege's logic aims at providing the material necessary for understanding the references to that logic given in the article. It is neither intended as a presentation of Frege's often subtle thoughts, nor as a presentation of the number theories for which Frege developed his logic, nor as a presentation of Frege's terminology and notations that have become outdated. 
This brief introduction owes to both Franz von Kutschera \cite{Kutschera} and John P.\ Burgess \cite{Burgess} even though it slightly differs from the presentations of Frege's logic by these authors. 

Frege's logic is defined in three books: 
\begin{itemize}
   \item ``Begriffsschrift, eine der arithmetischen nachgebildete Formelsprache des reinen Denkens'' \cite{begriffsschrift} translated in English as ``A Concept Notation: A formula language of pure thought, modelled upon that of arithmetic'' \cite{Concept-Notation}

   \item ``Grundgesetze der Arithmetik'' volumes I and II \cite{Grundgesetze-der-Arithmetik-I,Grundgesetze-der-Arithmetik-II} partly translated in English as ``The Basic Laws of Arithmetic'' \cite{Basic-Laws-of-Arithmetic}. 
\end{itemize}
Frege's logic is the archetype of predicate logic as it is known today. However, it departs from predicate logic in several aspects of various significance. 

A first salient but insignificant aspect of Frege's logic is its two-dimensional syntax that, even though decried by logicians, makes much sense to a computer scientist because it reminds of how, since the 70es of the 20th century, structured programs \cite{boehm-jacopini-theorem,goto-statement-considered-harmful} are rendered, or ``pretty printed'', for better readability. 

A second salient and important aspect of Frege's logic is that its syntax covers both what are called today the logical and (some of) the meta-logical language. In a manner that reminds of Prolog meta-programming, Frege's logic language includes notations for assumptions, theorems, logical equivalence (noted nowadays $\models\hspace{-0.19em}\rmodels$), and the extension of a predicate (in the sense of the assignment of truth values to atoms), and the truth values ``true'' and ``false''. 
Frege's logic includes the following symbols that, nowadays, are seen as meta-logical: 
\begin{itemize}
  \item $=$ used for expressing that its two arguments (sentences) have the same truth value (depending on the context, this is expressed nowadays using $\models\hspace{-0.19em}\rmodels$ or $\Leftrightarrow$)
  \item $\mid\hspace{-0.4em}-$ used for introducing an assumption
  \item $\mid\mid\hspace{-0.4em}-$ used for introducing a theorem
  \item $\text{ext}$ used as in $\text{ext}~ \alpha~ \Phi(\alpha)$ for denoting the ``course of values'' or, as the notation suggests, the extension, that is, the graph of the function $\Phi$
\end{itemize} 
Frege's logic also includes the symbol   --  as a mark for the beginning of a (two-dimen\-sional) sentence. 

A third salient aspect of Frege's logic is that the denotation, or literal meaning, of each of its sentences (or closed formulas) is a truth value ``true'' or ``false''. Consistently with this, predicates called ``concepts'' in Frege's logic are functions mapping their defining sentences to truth values. For avoiding confusions, the name ``concept'' is used in the following for referring to the functional predicates of Frege's logic. 

Frege's logic has first-order terms examples of which are the numbers $1$ and $3$ and the compound term $1+3$ built using the function symbol $+$. 
Frege's logic has atomic sentences built from concepts of arity 1 or 2. Interestingly, Frege's logic has no concepts of arity 0 that would correspond to propositional variables. Frege did not make use of concepts of arities greater than 2 because his number theories are conveniently expressed without such concepts. 
In Frege's logic, compound sentences are built very much like in nowadays' predicate logic from the connectives $\neg$ and $\Rightarrow$ and the universal quantifier $\forall$ that can be applied to terms as well as concepts.
As Frege points out, concepts can be first-order (if they predicate only of terms) or second-order (if they predicate of concepts): 
\begin{quote}
``We call first-order functions functions the arguments of which are objects, second-order functions functions the arguments of which are first-order functions.'' \cite[p.~36]{Grundgesetze-der-Arithmetik-I}\footnote{``Wir nennen nun die Functionen, deren Argumente Gegenst{\"a}nde sind, Functionen erster Stufe; die Functionen dagegen, deren Argumente Functionen erster Stufe sind, m{\"o}gen Functionen zweiter Stufe heissen.'' }
\end{quote}

Frege points out how conjunction ($\wedge$), disjunction ($\vee$), and equivalence ($\Leftrightarrow$) can be expressed using negation ($\neg$) and implication ($\Rightarrow$) and how existential quantification ($\exists$) can be expressed using negation ($\neg$) and universal quantification ($\forall$). 

Frege's logic has a proof calculus but no concept of interpretation, that is, no model theory. Interpretations would be introduced later and used in 1930 by G{\"o}del for proving the completeness of the proof calculus of Frege's logic \cite{goedel-thesis-german}. 

In Frege's logic, concept symbols but no formulas can occur as argument of second-order concepts. 
Let $S[a]$ denote a sentence in which the constant $a$ might occur and let $S[x/a]$ denote the expression obtained by replacing in $S$ each occurrence of $a$ by a variable $x$. Basic Law V of Frege's logic is an axiom  making it possible to define a concept $c$ from a sentence $S$ by an expression amounting to $\forall x~ c(x) = S[x/a]$. Basic Law V  \cite[$\S$ 20 p.~35]{Grundgesetze-der-Arithmetik-I} states:
\[\mid\hspace{-0.4em}-~ (\text{ext}~ \epsilon~ (f(\epsilon)) = \text{ext}~ \alpha~ (g(\alpha))) = \forall \textfrak{a}~ (f(\textfrak{a}) = g(\textfrak{a}))\] 
which means that the courses of values of two concepts, $\epsilon$ and $\alpha$, are identical if and only if the open formulas  defining these concepts, $f$ and $g$ respectively, have the same truth values for all the values of their variables. (Note the use in Basic Law V of the symbol $\mid\hspace{-0.4em}-$ for ``assumption'' or ``axiom'', and the second and third occurrences of $=$ expressing that two formulas denote the same truth value, that is, are logically equivalent.)

Basic Law V makes it possible to define what Frege calls a ``first-order concept'' $c$ that holds of all natural numbers (that is, non-negative integers) that are both even and odd. Since $c$'s defining expression, natural numbers being both even and odd, is inconsistent, the course of values $\text{ext}(c)$ of $c$ is, in nowadays notation, $\{(n, false) \mid n \in \mathbb{N}\}$ and $c$ specifies an empty set. 

Basic Law V also makes it possible to define what Frege calls a ``second-order concept'' $e$ as follows:
\begin{lstlisting}
   $(\star\star)~ \mid\hspace{-0.4em}-~ \forall x~ e(x) = \neg x(x)$ 
\end{lstlisting}
that is, apart from the use of symbol $\mid\hspace{-0.4em}-$ for expressing an assumption  and of $=$ instead of $\Leftrightarrow$, is exactly the definition of Russell's Paradox in Reflective Predicate Logic ($\star$) given in Section \ref{sec:Paradoxes}. 

In contrast to the aforementioned concept $c$ that holds of all natural numbers that are both even and odd, $e$'s defining expression $\neg x(x)$, a concept not holding of itself, is not inconsistent. It is the whole sentence $\forall x~ e(x) = \neg x(x)$ that is inconsistent. Indeed, instantiating the variable $x$ with $e$ in that sentence  yields $e(e) = \neg e(e)$, that is, in nowadays syntax $(e(e) \Leftrightarrow \neg e(e))$. Since its tentative definition is an inconsistent sentence, $e$ does not exist, hence its course of values does not exist either, that is, $e$ does not specify anything at all, not even an empty set. 

Russell's Paradox reminds of a propositional variable $p$ defined by the sentence $p = \neg p$ in Frege logic's syntax, or $(p \Leftrightarrow \neg p)$  in nowadays syntax. Since its tentative definition is an inconsistent sentence, $p$ does not exist, hence its course of values does not exist either, that is, $p$ does not specify anything at all. 

Thus, the inconsistency of Frege's logic does not result from Basic Law V in itself. It results from both, Frege logic's impredicative atoms and Basic Law V, that together make possible inconsistent concept definitions like Russell's paradox $(\star\star)$. 

Russell devised his Ramified Theory of Types \cite{ramified-theory-of-types} so as to make impossible inconsistent sentences like that of the paradox bearing his name. Russell's Ramified Theory of Types requires that a higher-order concept applies only of concepts of the immediately preceding order. In other words, Russell's Ramified Theory of Types precludes impredicative atoms like $e(e)$ or \lstinline{belief(belief(itRains))}. Thus, Russell's fix of Frege's logic ensures the consistency of axioms resulting from Basic Law V by precluding not only inconsistent axioms but also some expressions including consistent sentences. 

Interestingly, the  Ramified Theory of Types does not preclude that a propositional variable $p$ is defined by the inconsistent sentence $(p \Leftrightarrow \neg p)$. 
Russell's position was not flawed, though. Since the Vicious Circle Principle  \cite{russell-vicious-circle-principle-1,russell-vicious-circle-principle-2} he advocated for (see Section\ref{sec:Predicativity-Impredicativity}) forbids to define something by referring to that same thing, that principle also forbids to define a propositional variable $p$ by $(p \Leftrightarrow \neg p)$. Thus, Russell had no reasons to preclude such inconsistent definitions of propositional variables with his Ramified Theory of Types. 
 
It is nonetheless puzzling that the obvious fix of Frege's logic consisting in requiring that concept definitions are consistent would have been immediately apparent if Frege had included propositional variables in his logic that, three decades earlier, George Boole had proposed and formalised \cite{boole-propositional-logic}. 

It is tempting to think that following Frege's inspiration, a logic with impredicative atoms similar to Reflective Predicate Logic could have been proposed and accepted much earlier. This, however, is doubtful. Indeed, one essential step towards Frege's goal of specifying number theories was to provide a set theory. Russell's objection to Frege's logic was rooted at the kind of ``collections'' Frege's logic and Reflective Predicate Logic, because of their impredicative atoms, give rise to define. Such ``collections'', like the collection of beliefs mentioned in Section \ref{sec:Predicativity-Impredicativity}, make much sense in knowledge representation and in formalising meta-programming. For specifying number theories, however, they are more complicated than necessary. 

\bibliography{bry-in-praise-3rd-final-revision}

\label{lastpage}
\end{document}